\documentclass[reprint,superscriptaddress,amsmath,amssymb,aps,prb,floatfix,english,]{revtex4-2}

\usepackage{bibunits}
\usepackage[unicode=true, colorlinks=true, citecolor={blue!80!black}, urlcolor={blue!50!black}, linkcolor = {blue!80!black}]{hyperref}
\defaultbibliography{references}
\defaultbibliographystyle{apsrev4-2}
\usepackage{graphicx}
\usepackage{svg}
\usepackage{dcolumn}
\usepackage{siunitx}
\usepackage[T1]{fontenc}
\usepackage[utf8]{inputenc}
\usepackage{bbm}
\usepackage{array}
\usepackage{braket}
\usepackage{multirow}
\usepackage{mathtools}
\usepackage{comment}
\usepackage{footmisc}
\usepackage{hyperref}

\newcommand{\mK}{\:\mathrm{mK}}
\newcommand{\kb}{\:k_\mathrm{B}}
\newcommand{\xqp}{\:x_{\mathrm{QP}}}
\newcommand{\dD}{\:\delta\Delta}

\newcommand{\Gqp}{\:\Gamma^\mathrm{QP}}
\newcommand{\gqp}{\:\gamma^\mathrm{QP}}

\newcommand{\hfq}{\:hf_q}
\newcommand{\xqpne}{\:x_{\mathrm{QP}}^\mathrm{ne}}

\newcommand{\exponent}{\:(\delta\Delta-hf_q)/k_\mathrm{B} T}

\begin{document}

\preprint{APS/123-QED}

\title{Recovery dynamics of a gap-engineered transmon after a quasiparticle burst}
\author{Heekun~Nho}
\thanks{heekun.nho@yale.edu}
\affiliation{Departments of Applied Physics and Physics, Yale University, New Haven, CT 06520, USA}
\author{Thomas~Connolly}
\affiliation{Departments of Applied Physics and Physics, Yale University, New Haven, CT 06520, USA}
\author{Pavel~D.~Kurilovich}
\affiliation{Departments of Applied Physics and Physics, Yale University, New Haven, CT 06520, USA}
\author{Spencer~Diamond}
\thanks{{Present address: Northrop Grumman Corporation, Linthicum, MD 21090, USA}}
\affiliation{Departments of Applied Physics and Physics, Yale University, New Haven, CT 06520, USA}

\author{Charlotte~G.~L.~B\o ttcher}
\thanks{{Present address: Department of Applied Physics, Stanford University, Stanford, CA 94305, USA}}
\affiliation{Departments of Applied Physics and Physics, Yale University, New Haven, CT 06520, USA}
\author{Leonid~I.~Glazman}
\affiliation{Departments of Applied Physics and Physics, Yale University, New Haven, CT 06520, USA}
\affiliation{Yale Quantum Institute, Yale University, New Haven, CT 06511, USA}
\author{Michel~H.~Devoret}\thanks{michel.devoret@yale.edu\\
{Present address: Physics Dept., U.C. Santa Barbara, Santa Barbara, California 93106, USA and Google Quantum AI, 301 Mentor Dr, Goleta, CA 93111, USA}}
\affiliation{Departments of Applied Physics and Physics, Yale University, New Haven, CT 06520, USA}

\date{\today}

\begin{abstract}
Ionizing radiation impacts create bursts of quasiparticle density in superconducting qubits. These bursts temporarily degrade qubit coherence which can be detrimental for quantum error correction. Here, we experimentally resolve quasiparticle bursts in 3D gap-engineered transmon qubits by continuously monitoring qubit transitions. Gap engineering allows us to reduce the burst detection rate by a factor of five. This reduction falls four orders of magnitude short of that expected if the quasiparticles were to quickly thermalize to the cryostat temperature. We associate the limited effect of gap engineering with the slow thermalization of the phonons in our chips after the burst.
\end{abstract}

\maketitle

\textit{Introduction.---}Quantum error correction protocols redundantly encode a logical qubit in the entangled state of multiple physical qubits \cite{nielsen_quantum_2010}, making encoded information robust to errors on individual qubits. This holds if errors are uncorrelated between different physical qubits and lack temporal correlations on the same qubit.
Common errors in superconducting processors (e.g. qubit relaxation from material defects) generally satisfy these assumptions. However, there are rare error burst events that violate them.
During these error bursts (hereby referred to as ``bursts''), qubit relaxation rates temporarily increase across multiple qubits simultaneously \cite{vepsalainen_impact_2020,martinis_saving_2021,wilen_correlated_2021,mcewen_resolving_2022,thorbeck_two-level-system_2023,harrington_synchronous_2025,li_cosmic-ray-induced_2025}. Mitigating bursts is therefore essential to realize large-scale quantum error correction.

The underlying mechanism of the bursts is the impact of ionizing radiation on the chip hosting the qubits \footnote{There is another mechanism for bursts, where a phonon shower is generated by the avalanche-like release of stress in the chip \cite{yelton_correlated_2025}. Its experimental signatures and mitigation strategies should be similar to those for high-energy impacts.}. As illustrated in Fig.~\ref{fig:fig2}(a), a high-energy particle striking the chip produces a cascade of high-energy phonons that rapidly propagate throughout the substrate. When the phonons collide with superconducting films of the qubits, they break Cooper pairs, producing quasiparticle excitations (QPs). When the QPs tunnel across the Josephson junction, they can exchange energy with the qubit degree of freedom. This results in an elevated qubit relaxation rate. Only when the QPs recombine or get trapped does the error burst end and the processor performance recover. This prompts the search for strategies to either accelerate the phonon decay \cite{martinis_saving_2021,iaia_phonon_2022} or suppress QP tunneling.

On the latter front, a promising way forward is engineering of the superconducting gap \cite{aumentado_nonequilibrium_2004,ferguson_microsecond_2006,naaman_narrow-band_2007,court_energy_2007,shaw_kinetics_2008,sun_measurements_2012,riwar_efficient_2019,serniak_nonequilibrium_2019,diamond_distinguishing_2022,pan_engineering_2022,marchegiani_quasiparticles_2022,connolly_coexistence_2024,mcewen_resisting_2024}. 
A non-uniform gap profile, which can be created by varying superconducting film thicknesses, can impede QP tunneling across the qubit junction and suppress decoherence \cite{diamond_distinguishing_2022,marchegiani_quasiparticles_2022,connolly_coexistence_2024,mcewen_resisting_2024}. 
For this approach to be effective, the difference in gaps at the two sides of the junction should exceed the qubit energy by an amount significantly larger than the QP energy. 
This condition has been shown to hold for the ``resident'' non-equilibrium QPs typically present in the superconducting films in the steady state \cite{connolly_coexistence_2024}. By interacting with phonons, the energy of these QPs equilibrates with the base plate of the refrigerator even though their density remains strongly out of equilibrium. Yet, an important question for the suppression of QP-induced correlated errors remains unanswered: can gap engineering prevent QP tunneling following the impact of high-energy particles, when the phonon bath itself is moved away from equilibrium?

In this paper, we address this question by measuring QP bursts in transmons with different pairs of junction film thicknesses and, consequently, with different gap profiles. All of our transmons are made of aluminum, have 3D geometry \cite{paik_observation_2011}, and are deposited on sapphire substrates.
First, we benchmark our gap engineering by measuring the gap difference, $\delta\Delta$, for our devices with different film thicknesses. The gap difference is obtained by monitoring the tunneling rate of resident QPs at different fridge temperatures. By varying the film thickness, we control the gap difference from 0 to 10 $\mathrm{GHz}$ (in frequency units). We then evaluate the effect of gap difference on the detection rate of burst events and measure qubit relaxation rates during bursts. 

We find that gap engineering reduces the burst detection rate by a factor of five. However, the reduction falls short of the four orders of magnitude suppression observed for the resident QPs. We believe that this discrepancy results from the elevation of the effective chip temperature during the burst to about 90 mK. The additional thermal energy allows the QPs to overcome the gap difference and tunnel across the junction, thus facilitating qubit relaxation. We directly confirm the persistence of elevated substrate temperature by comparing the qubit excitation and relaxation rates during the burst.

\begin{figure*}[tbh]
  \begin{center}
    \includegraphics[scale = 1]{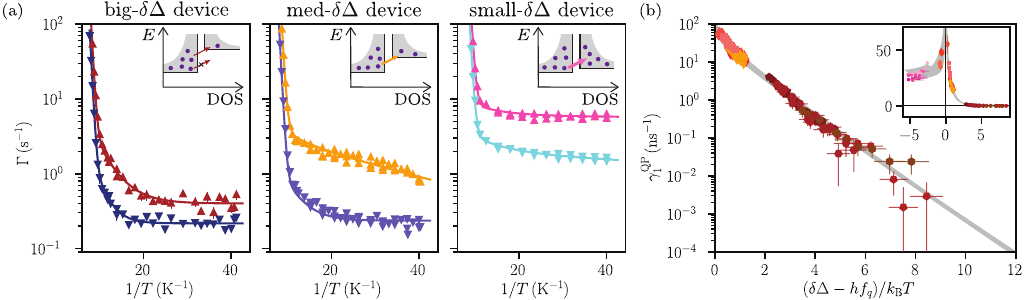}
\caption{ \label{fig:fig1}
(a)	Parity-switching rate ($\Gamma$) as a function of the fridge temperature ($T$) for devices with different film thicknesses. The film thickness is used as a knob to control the gap difference~$\dD$ at the qubit junction. Down (up)-pointing triangles represent rate $\Gamma_0$ ($\Gamma_1$) conditioned on the qubit residing in the ground (excited) state. When $\dD > hf_q$, the gap difference is efficient in obstructing the tunneling of resident QPs. Correspondingly, $\Gamma_1$ follows an Arrhenius law with the activation exponent $\dD - hf_q$, i.e., $\Gamma_1\propto \exp\left(-\left[\dD - hf_q\right] / \kb T\right)$. This regime is realized in the big- and medium-$\dD$ device (left and middle panels).
For the small-$\dD$ device (right panel), $\dD < hf_q$, and the gap difference is not effective at preventing the tunneling of resident QPs. The observed rate $\Gamma_1$ for the small-$\dD$ device is the highest among the measured devices. Comparing the data for $\Gamma_1$ and $\Gamma_0$ to theory (solid lines) allows us to extract $\dD$ and the resident QP density for all three devices. The resulting fit parameters are shown in Table~\ref{tab:table1}. Notably, $\Gamma^\mathrm{ph}$ and $\xqpne$ for the three devices are of the same order of magnitude despite being fabricated individually. We attribute this consistency to the shared electromagnetic and radiation environment provided by the multiplexed package \cite{noauthor_see_nodate}.
(b) Exponential suppression of the resident QP tunneling rate, $\Gamma_1-\Gamma_1^\mathrm{ph}$, in state $|1\rangle$. Different points correspond to devices with different $\dD-\hfq$, different cooldowns, and temperatures \cite{noauthor_see_nodate}. 
For convenience of comparison between different devices, the rate is normalized by the resident QP density, $\gqp_1\coloneqq (\Gamma_1-\Gamma_1^\mathrm{ph})/\xqpne$. When the gap difference surpasses the qubit energy, the normalized tunneling rate decreases exponentially with increasing $\exponent$. The inset shows the full dependence of $\gqp_1$ on $\exponent$. The normalized rate diverges when $\dD=\hfq$ but becomes temperature-insensitive for $\dD<\hfq$. The gray-shaded regions in the main plot and inset represent the theoretical prediction for the rate for a range of additional parameters such as $E_J$ or $E_C$.
}
  \end{center}
\end{figure*}
Our results should be contrasted with Ref.~\cite{mcewen_resisting_2024}, where the gap difference -- obtained by film thicknesses similar to our gap-engineered device -- resulted in essentially no bursts being detected.
The difference may arise from a combination of several factors. First, unlike ours, the chip in Ref.~\cite{mcewen_resisting_2024} has most of its area covered by the aluminum ground plane. Second, their substrate is silicon whereas ours is sapphire. Third, the chips are thermalized differently. Our chips are anchored to the package only by two isolated clamps each \cite{noauthor_see_nodate}, whereas Ref.~\cite{mcewen_resisting_2024} presumably used numerous wirebonds that anchor their chip. Finally, unlike in our measurement, Ref.~\cite{mcewen_resisting_2024} probed the qubit only sparsely, once every 100 $\mathrm{\mu s}$. Such a measurement may miss bursts recovering faster than 100 $\mathrm{\mu s}$.

\textit{Experimental setup.---}
Our goal is to evaluate the effectiveness of gap engineering in suppressing decoherence caused by the QP bursts. The gap difference between the junction leads, with gaps of $\Delta$ and $\Delta+\dD$, prevents the tunneling of low-energy QPs from the low-gap lead to the high-gap one. The elevation of the superconducting gap in one of the leads, $\delta \Delta$, is controlled by changing the film thicknesses employed in the junction fabrication \cite{chubov_dependence_1969, yamamoto_parity_2006, court_energy_2007, diamond_distinguishing_2022, marchegiani_quasiparticles_2022, connolly_coexistence_2024}. We fabricate three 3D aluminum transmons using different film thickness pairs: [17 nm, 83 nm], [23 nm, 123 nm], and [85 nm, 106 nm] (measured by atomic force microscopy). Based on prior measurements \cite{chubov_dependence_1969, yamamoto_parity_2006,court_energy_2007,marchegiani_quasiparticles_2022}, the corresponding $\dD$ are expected to be approximately 10 GHz (``big''-$\dD$), 5 GHz (``medium''-$\dD$), and 0.5 GHz (``small''-$\dD$), respectively, in frequency units. Table~\ref{tab:table1} summarizes relevant qubit parameters.

The transmons are designed to be in the offset-charge-sensitive regime \cite{riste_millisecond_2013,serniak_hot_2018,serniak_direct_2019}, with the charge dispersion on the order of MHz. This enables the detection of charge-parity-switching events, which are experimental signatures of QP tunneling. By measuring the parity-switching rates, we can monitor how frequently QPs cross the junction in the presence of $\dD$. All three chips are mounted together in a multiplexed package, ensuring that electromagnetic environment is the same \cite{noauthor_see_nodate}.

\begin{center}
\begin{table}[tbh]
\caption{\label{tab:table1}
Fit parameters and uncertainties corresponding to devices in Fig.~\ref{fig:fig1}(a). 
For the medium- and small-$\dD$ devices, the parity-switching rate is dominated by the QP tunneling rate, making it difficult to extract $\Gamma_1^\mathrm{ph}$. Therefore, for these devices, we assume $\Gamma_0^\mathrm{ph} = \Gamma_1^\mathrm{ph}$. In contrast, we can confidently fit $\Gamma_1^\mathrm{ph}$ for the big-$\dD$ device. This value can be found in \cite{noauthor_see_nodate}.}
\newcolumntype{P}[1]{>{\centering\arraybackslash}p{#1}}
\begin{tabular}{c|c|c|c|c|c}
\hline\hline
  & $f_q$ & $\dD/h$ & $\Delta/h$ & $\xqpne $ & $\Gamma_0^\mathrm{ph}$   \\ 
  & (GHz) & (GHz)   &  (GHz)  & ($\times 10^{-10}$) &  ($\mathrm{s}^{-1}$) \\
  \hline
 big-$\dD$ & 3.57 & 8.2$\pm$0.4 & 45.2$\pm$0.4 & 6.3$\pm$1.8 & 0.22$\pm$0.01\\  

 \hline
 med.-$\dD$ & 4.78 & 5.8$\pm$0.1 & 46.1$\pm$0.1 & 1.0$\pm$0.1 & 0.24$\pm$0.01 \\
 \hline
 small-$\dD$ & 4.27 & 0$\pm$1.1 & 45.1$\pm$0.4 & 1.9$\pm$0.1 & 0.52$\pm$0.05 \\
 \hline\hline

\end{tabular}
\end{table}
\end{center}

\textit{Measuring the gap difference.---}As an initial step, we characterize the superconducting gap barriers, $\delta\Delta$, of our devices. To do this, we measure the parity-switching rates in the steady state, i.e., away from rare QP bursts. In this case, the parity-switching rates are dominated by the tunneling of resident non-equilibrium QPs populating superconducting films. The energy distribution of these QPs equilibrates with the base stage of the refrigerator \cite{connolly_coexistence_2024}. The temperature dependence of the QP tunneling rate then follows the Arrhenius law, with the activation exponent controlled by $\dD$. Indeed, $\dD$ impedes the QPs across the junction, such that only few energetic QPs tunnel through the junction. Fitting the measured temperature dependence of the parity-switching rates to the Arrhenius law thus directly yields $\dD$, as we explain below. This measurement resembles that in Ref.~\cite{connolly_coexistence_2024}.
Fig.~\ref{fig:fig1}(a) shows the measured parity-switching rates as a function of temperature for our three devices, and Fig.~\ref{fig:fig1}(b) shows a summary of additional devices and cooldowns.

To fit the measured rates, we use a detailed theoretical model of QP tunneling in the presence of a gap difference at the junction \cite{diamond_distinguishing_2022,marchegiani_quasiparticles_2022,connolly_coexistence_2024}. According to this theory, when the qubit is in the ground state, the parity-switching rate reads $\Gqp_0 = \gqp_0 \xqpne$. Here, $\xqpne$ is the temperature-independent density of resident QPs (in units of Cooper pair density), and $\gqp_0 \propto E_J \exp\left(-\dD / \kb T\right)$ 
is proportional to the tunneling rate and the fraction of QPs with energy exceeding $\delta\Delta$. Here, $\kb$ is the Boltzmann constant, $T$ is the QP temperature, and $E_J$ is the Josephson energy of the junction.
When the qubit is excited, and if $\dD>\hfq$, this expression modifies to $\Gqp_1 = \gqp_1 \xqpne$ with $\gqp_1 \propto E_J \exp\left(-\left[\dD - \hfq\right]/ \kb T\right)$, where $h$ is the Planck constant and $f_q$ is the qubit frequency. The activation energy is reduced compared to that in $\gqp_0$ because the QP can absorb the excitation from the qubit, making it easier to overcome the gap difference.
On contrary, if $\dD<\hfq$, then $\gqp_1$ becomes roughly temperature-independent. For each device, we simultaneously fit $\Gamma_0$ and $\Gamma_1$ to the expressions outlined above, using $\dD$ and $\xqpne$ as fitting parameters.

There are two complications in this fitting protocol. First, at high temperatures, $T> 100\mK$, the density of equilibrium QPs,  $\xqp^\mathrm{th}(T)=\sqrt{2\pi\kb T/\Delta}\exp\left(-\Delta / \kb T\right)$, becomes comparable to that of the resident non-equilibrium QPs, $\xqpne$. Therefore, at high temperatures, we account for the contribution of equilibrium QPs to the parity-switching rates. Second, there is a small background contribution $\Gamma^\mathrm{ph}$ to parity-switching rates coming from the absorption of stray pair-breaking photons at the junction \cite{houzet_photon-assisted_2019,diamond_distinguishing_2022,connolly_coexistence_2024}. In our theoretical fit, we add such contributions as a temperature-independent offset. Notably, the contribution $\Gamma^\mathrm{ph}$ of the photons might differ for the two qubit states \cite{houzet_photon-assisted_2019,diamond_distinguishing_2022}. We use notations $\Gamma_0^\mathrm{ph}$ and $\Gamma_1^\mathrm{ph}$ to differentiate this contribution between the respective qubit states.

\begin{figure*}[t]
  \begin{center}
    \includegraphics[scale = 1]{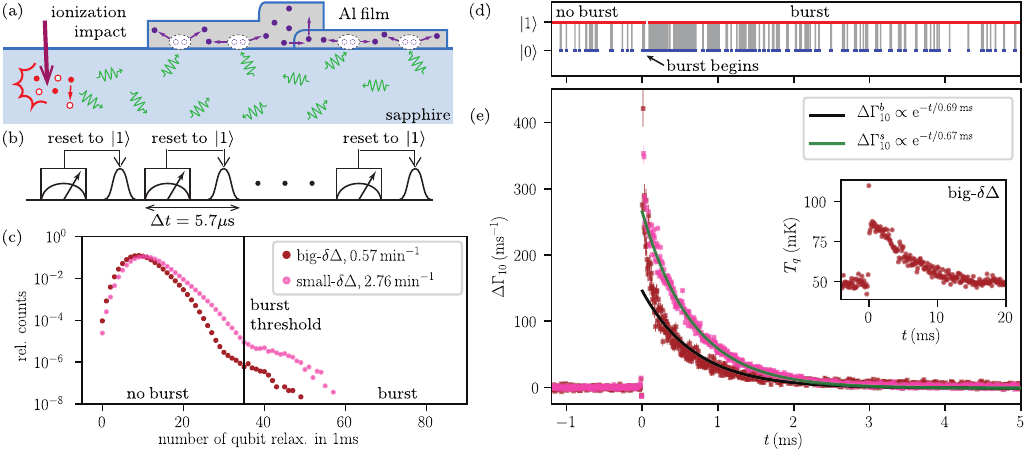}
\caption{ 
(a)	Schematic of how a high-energy impact leads to a QP burst. The collision of high-energy particles ionizes electron-hole pairs in the substrate. When the electron-hole pairs recombine, they generate high-energy phonons (green waves). These phonons have enough energy to break Cooper pairs in superconducting films, thereby generating numerous QPs. Tunneling of these QPs across the junction increases the qubit relaxation rate.
(b)	Pulse sequence for detecting QP bursts. Every 5.7 $\mathrm{\mu s}$, the qubit is read-out, then projected into $\ket{1}$. 
(c)	Histogram of the qubit relaxation event counts within a 1 ms window. Different colors correspond to different devices. In the absence of bursts, we expect the histograms to represent a Poisson process. The most likely number of relaxation events is determined by the steady-state coherence time $T_1^{\rm steady}$. The positions of the observed maxima are consistent with measured steady-state relaxation times ($T_1^{\rm steady} \approx 100 \mathrm{\mu s}$).
We associate clear deviation from the Poissonian statistics in the tails of the distributions with the rare QP bursts.
To the right of the black line, the measured counts for both devices are dominated by QP bursts. Post-selecting events based on this threshold thus allows fair comparison of the burst occurrence rate in the big-$\dD$ and small-$\dD$ devices. The larger $\dD$ yields roughly five times fewer detected burst events according to this criteria.
(d)	A readout trace during a QP burst. After the burst begins, the qubit experiences an excess number of relaxation events. During the burst, the number of relaxation events in several 1 ms bins exceeds the threshold in panel (c). It allows us to identify the burst.
(e)	Excess qubit relaxation rate $\Delta\Gamma_{10}$ after high-energy impacts, averaged over all observed burst events. The bursts are identified according to the threshold defined in panel (c). 
Exponential fit results are shown with solid lines. The extracted burst duration, see the legend, is insensitive to $\dD$.
The inset shows the qubit temperature, $T_q$, during bursts in the big-$\dD$ device (see \cite{noauthor_see_nodate} for the measurement protocol). We observe $T_q$ abruptly increases from the background value (50 mK) to $\approx90$ mK and stays consistently elevated for over 5 ms. We attribute this increase to elevated QP temperature during the burst. We note that the 50 mK background likely does not reflect the real temperature of the sample. Excess of $\ket{1}$ measurements primarily stems from measurement errors.
}\label{fig:fig2}
  \end{center}
\end{figure*}
\textit{Burst measurement.---}With the gap differences under control, we proceed to analyze the temporally correlated errors caused by QP bursts in the devices. To assess the effect of the gap difference on mitigating the high-energy impacts, we compare the bursts in the big- and small-$\dD$ devices. The results for the medium-$\dD$ device are discussed separately in \cite{noauthor_see_nodate}, due to the complications arising from its gap difference being comparable to the qubit energy.

To detect the QP bursts, we attempt to stabilize the qubit in its excited state \cite{mcewen_resolving_2022,harrington_synchronous_2025,dominicis_evaluating_2024,wu_mitigating_2025}. To this end, we repeatedly readout the qubit state every 5.7 $\mathrm{\mu s}$. If we find the qubit in its ground state $|0\rangle$, we apply a $\pi$-pulse to bring it to the excited state $|1\rangle$. The pulse sequence is illustrated in Fig.~\ref{fig:fig2}(b). During QP burst events, the qubit relaxation rate significantly increases. This allows us to identify the bursts by the excess number of observed $|1\rangle \rightarrow |0\rangle$ relaxation events [e.g. Fig.~\ref{fig:fig2}(d)]. In our measurement, we post-select events that do not cause large changes to the offset charge.
However, we find that this post-selection is inconsequential for the burst event statistics \cite{noauthor_see_nodate}.

To collect sufficient statistics, we repeat the measurement for about 10 hours for each device. We then chop the measurement traces into $1\:\mathrm{ms}$ windows and tally the number of observed relaxation events in each window. Normalized histograms of the relaxation counts are shown in Fig.~\ref{fig:fig2}(c). For each device, the histogram contains a large peak (10 - 15 relaxation events), which is unrelated to the QP bursts. 
Although the peak of the distribution is described by the uncorrelated Poissonian relaxation events, there is an anomolous non-Poissonian tail containing an excess number of relaxation events. This tail stems from bursts containing correlated errors. 

Its clear distinction from the Poissonian peak makes us confident that a burst has occurred within a $1\:\mathrm{ms}$ window when the number of relaxation events exceeded a certain threshold. 
In practice, we set the threshold at eight standard deviations away from the mean of both Poissonian distributions to make the false positive event rate negligible.
Notably, some of the identified events are unrelated to bursts; they stem from unstable lossy defects in the environment \cite{muller_interacting_2015,klimov_fluctuations_2018,burnett_decoherence_2019,muller_towards_2019,schlor_correlating_2019,lisenfeld_electric_2019}. These events can be effectively filtered out since their time dynamics is sufficiently different from that of the bursts \cite{noauthor_see_nodate}.

Using the common threshold for both devices, we see that the burst detection rate in the big-$\delta\Delta$ device ($0.57\:\mathrm{min}^{-1}$) is lower than that in the small-$\dD$ device ($2.76\:\mathrm{min}^{-1}$).
Notably, the suppression of the burst detection rate by gap engineering is only about a factor of five. This observation is inconsistent with the QP energy thermalizing with the mixing chamber of the fridge immediately after the impact. If QPs were thermalized -- as for resident QPs in Fig.~1 -- $\dD$ would suppress the burst detection rate by about a factor of $\exp(\left[\dD - hf_q\right]/\kb T) \sim 10^4$ in the big-$\dD$ device. Here, $T = 25\:\mathrm{mK}$ is the base temperature of our fridge.

Next, we consider how $T_1$ of the qubit returns to its steady state after the impact. To this end, we use continuous monitoring data from the measurement of Fig.~\ref{fig:fig2}(b-d). We identify the bursts using the threshold shown in Fig.~\ref{fig:fig2}(c). Averaging the measurement outcomes between many observed burst events allows us to determine the qubit relaxation rate $\Gamma_{10}$ as a function of time after the impact \cite{noauthor_see_nodate}.
The results of this analysis are shown in Fig.~\ref{fig:fig2}(e). For both devices, the return of the relaxation rate to the steady state is rougly exponential. The associated time constants $\sim 0.7\:\mathrm{ms}$ are insensitive to the gap engineering.

We note that different ionizing radiation sources, such as muons or gamma rays, may produce bursts of varying severity since vastly different amounts of energy can be absorbed by the chip \cite{fowler_spectroscopic_2024}.
Our measurements lack resolution to distinguish between different ionizing radiation sources. This is the reason why we average the response of all bursts together, regardless of their severity or origin. In the supplementary materials, we provide evidence that up to an overall rescaling, the dynamics of the qubit's relaxation rate is independent of the burst severity \cite{noauthor_see_nodate}.
We leave a detailed investigation of different radiation sources to future work \cite{vepsalainen_impact_2020,harrington_synchronous_2025,li_cosmic-ray-induced_2025,fowler_spectroscopic_2024}.

In addition to the qubit \textit{relaxation} rate $\Gamma_{10}(t)$ after the impact, we similarly measure the qubit \textit{excitation} rate $\Gamma_{01}(t)$. To this end, we repeat the measurement similar to that described in Fig.~\ref{fig:fig2}(b), except without resetting the qubit to $\ket{1}$. From the two rates, one may extract the time dependence of the QP density, $\xqp(t)$, and the qubit temperature $T_q(t)$, see the inset of Fig.~\ref{fig:fig2}(e) for the latter quantity. Here, the qubit temperature is defined as $\kb T_q = hf_q / \mathrm{ln} \left(\Gamma_{10} / \Gamma_{01}\right)$. The time dependence of $\xqp$ is similar to that of $\Gamma_{10}(t)$. In about a millisecond, the QP density decays to a level where QPs no longer limit the qubit $T_1$ \cite{noauthor_see_nodate}.

The qubit temperature $T_q$ surges to about 90 mK after the impact, significantly exceeding the fridge's base temperature. The qubit temperature returns to its steady state exponentially, with a time constant of about $6\:\mathrm{ms}$. This time scale is almost 10 times longer than the $0.7\:\mathrm{ms}$ time constant for the decay of $\xqp$. The temperature thus remains elevated even long after the generated by the impact non-equilibrium QPs are gone. 
This points to a dominant role of the substrate properties in determining the evolution of $T_q(t)$.
Indeed, the observed increase in qubit temperature to $90\:\mathrm{mK}$ is consistent with the substrate phonon heat capacity for a typical radiation impact energy of $\sim 100\mathrm{keV}-1\:\mathrm{MeV}$ \cite{fowler_spectroscopic_2024,noauthor_see_nodate}.
The decay constant for the qubit temperature is determined by the rate of phonon escape from the system, likely through the chip clamps.

The elevated substrate temperature after burst explains the weak sensitivity of burst occurrence to the gap difference. As previously described, the suppression of QP-induced qubit relaxation by $\dD$ is controlled by the activation exponent $\exp(-[\delta\Delta - hf_q]/\kb T)$. For the gap difference to be effective, $\delta\Delta - hf_q$ must be much larger than $\kb T$. For our big-$\delta\Delta$ device, $(\delta\Delta - hf_q)/\kb \sim 200\:\mathrm{mK}$ which only exceeds the lattice temperature during the QP burst by a factor of two. This is consistent with the observed five-fold decrease of the burst occurrence in the presence of gap difference.

At short times ($t \lesssim 100 \:\mu\mathrm{s}$), $\Delta \Gamma_{10}(t)$ in the big-$\dD$ device deviates from exponential decrease, as shown in Fig.~\ref{fig:fig2}(e). We attribute this to the time dynamics of the QP distribution temperature. For $t \lesssim 100 \:\mu\mathrm{s}$, the effective qubit temperature $T_q$ spikes well above 100 mK before rapidly decaying to $\sim90$ mK for the remainder of the burst ($100 \:\mu\mathrm{s}<t<5\:\mathrm{ms}$) [Fig.~\ref{fig:fig2}(e) inset and Fig. S14(f) in supplementary materials]. This temperature spike is likely related to the abundance of high-energy phonons that quickly get down-converted \cite{kurilovich_correlated_2025}. Combined with the strong temperature dependence of $\gqp_{10}$ in big-$\delta\Delta$ device, the rapid spike and decay of the temperature is responsible for the non-exponential behavior of $\Delta \Gamma_{10}(t)$ during the initial part of the burst.

\textit{Conclusion.---}
In this work, we investigate the effect of superconducting gap engineering on QP bursts in transmon qubits. Our results confirm that the superconducting gap difference between the two sides of the junction exceeding the qubit energy, $\dD>hf_q$, can reduce both the burst detection rate, see Fig.~\ref{fig:fig2}(c), and the qubit relaxation rate during these events, see Fig.~\ref{fig:fig2}(e).

However, we also uncover a critical limitation of this method: during bursts, the substrate and QPs retain high energy $\sim 90\:\mathrm{mK}$, only mildly lower than $
\dD - hf_q$. We attribute this to the slow escape of hot phonons from our chip. Our findings suggest two potential strategies to further suppress the adverse effects of bursts. First, increasing the gap difference to $\dD/\kb\gg 100\:\mathrm{mK}$ can prevent QPs from crossing the junction even when their energy is elevated. The increase of $\dD$ can be achieved with high-gap materials, such as granular aluminum \cite{abeles_enhancement_1966,cohen_superconductivity_1968,deutscher_transition_1973,sun_measurements_2012}. Second, the effectiveness of gap engineering can be enhanced by providing an efficient escape mechanism for hot phonons. This involves improving the thermal link between the chip and the fridge's base plate. To achieve optimal thermalization, one should test various substrate materials and clamping schemes; this comparison is outside the scope of our work. Alternatively, the phonons can be absorbed by patches of normal metal placed directly on the chip \cite{martinis_saving_2021,iaia_phonon_2022,larson_quasiparticle_2025}.
Due to their high heat capacity, these patches can serve as efficient heat sinks.
Combining these strategies should suppress the detrimental effect of QP bursts on superconducting quantum processors.

\textit{Acknowledgements.---} We acknowledge Suhas Ganjam who provided package and qubit design. We acknowledge useful discussions with Valla Fatemi and Kaavya Sahay.
This research was sponsored by the Army Research Office (ARO) under grants no.W911NF-22-1- 0053 and W911NF-23-1-0051, by DARPA under grant no. HR0011-24-2-0346, by the U.S. Department of Energy (DoE), Office of Science, National Quantum Information Science Research Centers, Co-design Center for Quantum Advantage (C2QA) under contract number DE-SC0012704. The views and conclusions contained in this document are those of the authors and should not be interpreted as representing the official policies, either expressed or implied, of the ARO, DARPA, DoE, or the US Government. The US Government is authorized to reproduce and distribute reprints for Government purposes notwithstanding any copyright notation herein. Fabrication facilities use was supported by the Yale Institute for Nanoscience and Quantum Engineering (YINQE) and the Yale Univeristy Cleanroom.

\nocite{connolly_coexistence_2024,halpern_far_1986,rehammar_low-pass_2023,serniak_hot_2018,serniak_direct_2019,diamond_distinguishing_2022,houzet_photon-assisted_2019,barends_minimizing_2011,diamond_quasiparticles_2024,marchegiani_quasiparticles_2022,glazman_bogoliubov_2021,catelani_parity_2014,cherney_enhancement_1969,court_energy_2007,chubov_dependence_1969,nsanzineza_trapping_2014,vool_non-poissonian_2014,wang_measurement_2014,wilen_correlated_2021,fowler_spectroscopic_2024,fugate_specific_1969,viswanathan_heat_1975,yelton_correlated_2025}
\bibliography{references.bib}

\end{document}


\myexternaldocument{ms}
\beginsupplement

\title{Supplementary information for ``Recovery dynamics of a gap-engineered transmon after a quasiparticle burst''}
\author{Heekun~Nho}
\thanks{heekun.nho@yale.edu}
\affiliation{Departments of Applied Physics and Physics, Yale University, New Haven, CT 06520, USA}
\author{Thomas~Connolly}
\affiliation{Departments of Applied Physics and Physics, Yale University, New Haven, CT 06520, USA}
\author{Pavel~D.~Kurilovich}
\affiliation{Departments of Applied Physics and Physics, Yale University, New Haven, CT 06520, USA}
\author{Spencer~Diamond}
\thanks{{Present address: Northrop Grumman Corporation, Linthicum, MD 21090, USA}}
\affiliation{Departments of Applied Physics and Physics, Yale University, New Haven, CT 06520, USA}

\author{Charlotte~G.~L.~B\o ttcher}
\thanks{{Present address: Department of Applied Physics, Stanford University, Stanford, CA 94305, USA}}
\affiliation{Departments of Applied Physics and Physics, Yale University, New Haven, CT 06520, USA}
\author{Leonid~I.~Glazman}
\affiliation{Departments of Applied Physics and Physics, Yale University, New Haven, CT 06520, USA}
\affiliation{Yale Quantum Institute, Yale University, New Haven, CT 06511, USA}
\author{Michel~H.~Devoret}\thanks{michel.devoret@yale.edu\\
{Present address: Physics Dept., U.C. Santa Barbara, Santa Barbara, California 93106, USA and Google Quantum AI, 301 Mentor Dr, Goleta, CA 93111, USA}}
\affiliation{Departments of Applied Physics and Physics, Yale University, New Haven, CT 06520, USA}

\date{\today}

\maketitle

\tableofcontents

\vspace{2 cm}

\newpage

\section{Experimental setup}
\begin{figure*}[t]
  \begin{center}
    \includegraphics[scale = 1]{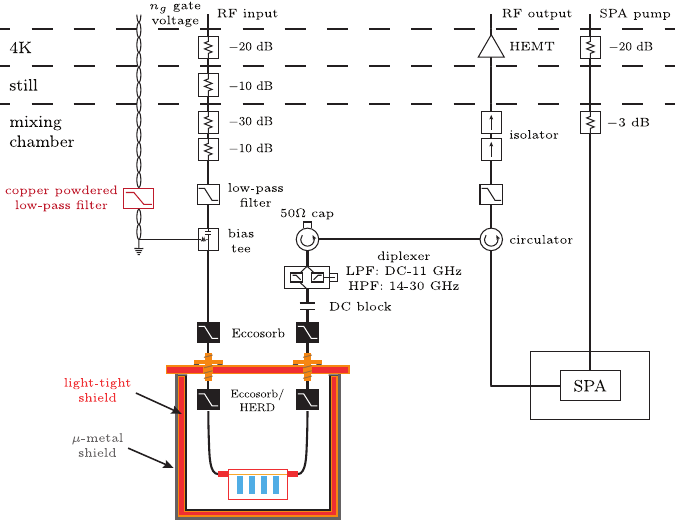}
\caption{ 
Wiring diagram of the measurement setup
}\label{fig:wiring}
  \end{center}
\end{figure*}

\subsection{Shielding and filtering against infrared photons}
\label{sec:shielding}
\begin{figure*}[h]
  \begin{center}
    \includegraphics[scale = 1]{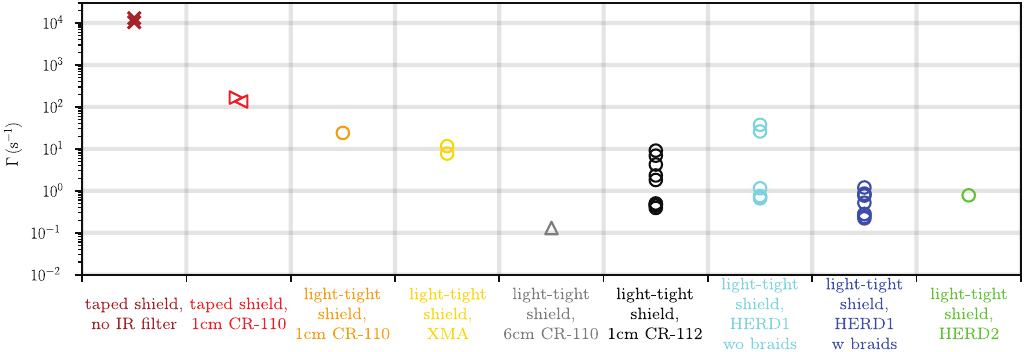}
\caption{ 
Parity-switching rate, $\Gamma$, of a transmon qubit with a different type of protection from infrared (IR) radiation. The protection consists of a shield enclosing the device and a filter (or multiple filters) in the readout line inside of the shield, see Fig.~\ref{fig:wiring} for an example. In a ``taped'' shield, the gaps exposed to IR radiation are covered with copper or aluminum tape. In a ``light-tight'' shield, the gaps are sealed with indium \cite{connolly_coexistence_2024}. 
For the filters, we use one of the following: (i) home-made filters where a segment of transmission line of a varied length is covered with Eccosorb CR-110 or Eccosorb CR-112 epoxy \cite{halpern_far_1986}; (ii) Eccosorb filter ordered from ``XMA'' company; (iii) HERD1 and HERD2 filters ordered from ``Quantum Microwave'' company~\cite{rehammar_low-pass_2023}. For HERD1, we differentiate the data based on whether it was additionally thermalized to the fridge's base plate using copper braids.
Non-circular data points are from previous works (X: Ref.~\cite{serniak_hot_2018}, \(\scalebox{1.3}{\(\triangleright\)}\,\): Ref.~\cite{serniak_direct_2019}, \(\scalebox{1.3}{\(\triangleleft\)}\,\): Ref.~\cite{diamond_distinguishing_2022}, $\triangle$: Ref.~\cite{connolly_coexistence_2024}). Circular data points are our measurements.
}\label{fig:sfig1}
  \end{center}
\end{figure*}
In our work, we often rely on the switching of charge parity in a transmon as a proxy for the tunneling of quasiparticles (QPs) across the junction. However, in addition to QP tunneling, the parity-switching can be caused by the absorption of Cooper-pair-breaking infrared photons at the junction \cite{houzet_photon-assisted_2019}. Therefore, to increase the sensitivity of our measurements to QP tunneling, it is crucial to suppress the infrared radiation incident on the device. Previous work, Ref.~\cite{connolly_coexistence_2024}, has shown that effective suppression of the infrared photon flux requires both device shielding and RF line filtering. 
Fig.~\ref{fig:wiring} illustrates the shielding and filtering configurations explored in our work, which we describe below.

For the shielding, we use a light-tight enclosure similar to the one described in Ref.~\cite{connolly_coexistence_2024}. To prevent the high-energy photons from leaking into the shield, we seal all seams -- such as gaps between the mezzanine plate and the shield or SMA connectors -- using indium O-rings. Additionally, we coat the inner wall of the shield can with an Stycast epoxy loaded with carbon powder to absorb IR photons which found their way into the shield \cite{barends_minimizing_2011,serniak_hot_2018}.
As long as the shield is effective, the only way for IR radiation to arrive at the sample is through the RF line (used for qubit readout and drive).

To filter out photons leaking through the RF line, we use two types of in-line IR filters in different configurations (see Fig.~\ref{fig:sfig1} for the summary): Eccosorb filter and Non-magnetic High-Energy Radiation-Drain Low pass filter (HERD). Our Eccosorb filter is custom-made. Its design consists of a transmission line surrounded by Eccosorb, an epoxy that absorbs high-energy photons. We use two different Eccosorb CR series, 110 and 112. The CR-112 produces stronger attenuation (12.5 dB insertion loss at 10 GHz per centimeter) than more commonly used CR 110 (1 dB at 10 GHz per centimeter). The HERD IR filters are produced by Sweden Quantum \cite{rehammar_low-pass_2023}. The HERD filter heavily attenuates signals above 80 GHz, but unlike Eccosorb has very low insertion loss at 10 GHz. It consists of a transmission line with many holes in the outer conductor. The holes allow high-frequency radiation to leak out and subsequenly get absorbed by a dissipative material.
Proper thermalization of the HERD1 filter is crucial. We once observed an elevated parity-switching rates by approximately two orders of magnitude (two outlier points in the cyan column of Fig.~\ref{fig:sfig1}). We conjecture that this might be attributed to the inadequate thermalization of the absorber part, as it was thermalized solely through SMA connectors. Notably, we never observe such a high photon-assisted parity-switching rate after we thermalize the outer absorber through copper braids (blue column of Fig.~\ref{fig:sfig1}).

Outside of the shield, we always employ Eccosorb filters in the RF readout line (CR-112 on the input and CR-110 on the output). Inside of the shield, we use either Eccosorb CR-112 filters or HERD filters, depending on the cooldown. We summarize the filter configurations used in different cooldowns in Table~\ref{tab:stable} and in Fig.~\ref{fig:sfig1}. For the data presented in the main text, inside of the shield, we use two HERD1 filters.

\subsection{Devices}
\begin{figure*}[h]
  \begin{center}
    \includegraphics[scale = 1]{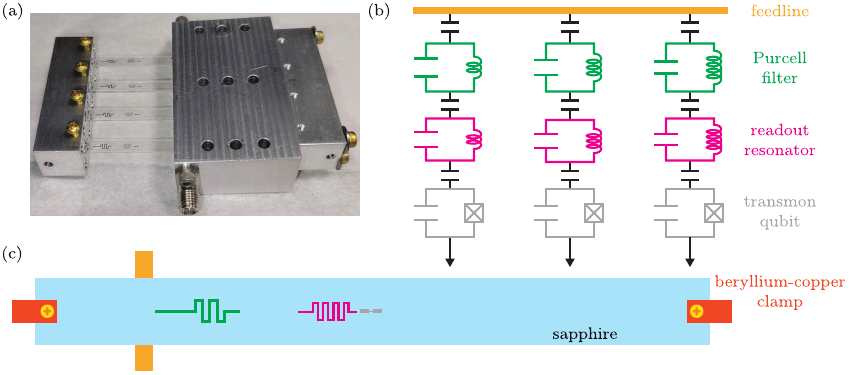}
\caption{
(a) Photograph of the package hosting the devices. In the photograph, the package is made out of aluminum. For the measurments presented in the main text, we use a similar package made out of copper. The devices are deposited on 4 cm $\times$ 4 mm $\times$ 0.5 mm sapphire chips, which are inserted into tunnels in the metal block. (b) Circuit diagram of the experiment. Three devices are capacitively coupled to the common feedline in a hanger configuration. 
Each device consists of a transmon qubit, a readout resonator, and a Purcell filter.
(c) Design of 3D transmon with a readout resonator and a Purcell filter. Both ends of the chip are clamped by beryllium-copper leaf springs.
}\label{fig:sfig2}
  \end{center}
\end{figure*}
Our devices are fabricated on a c-plane sapphire substrate and consist of 3D aluminum transmons, stripline readout resonators, and stripline Purcell filters, see Fig.~\ref{fig:sfig2}.  
The fabrication process follows that described in Ref.~\cite{serniak_direct_2019}, except the film thickness is varied during the double-angle deposition step. This variation is used to vary the superconducting gaps as we describe below. We summarize relevant device parameters in Table~\ref{tab:stable}.

The chips are mounted together in a multiplexed package, with up to four chips measured simultaneously during a single cooldown. Each device is capacitively coupled to a common feedline located under the chips in a hanger configuration. In the main text, we simultaneously measure three devices, see 2024.04 cooldown in Table~\ref{tab:stable}.

\section{Gap difference measurement and data analysis}
\label{sec:dD-msmt}
In our devices, we change the thickness of the films used in the Josephson junction fabrication. This allows us to create a gap difference at the junction which is intended to prevent the tunneling of QPs. This section details the characterization of the gap difference in our transmons, achieved by measuring the temperature dependence of the charge parity-switching rate. Section~\ref{sec:dD protocol} outlines the measurement protocol, which generally follows the methodology described in Ref.~\cite{connolly_coexistence_2024}. Section~\ref{sec:dD theory} presents the theoretical model used to fit the measured data. Section~\ref{sec:dD fit} describes the fitting procedure and the resulting fit. Section~\ref{sec:th_vs_D} investigates the correlation between the superconducting gap of aluminum films and their thickness.
\begin{figure*}[h]
  \begin{center}
    \includegraphics[scale = 1]{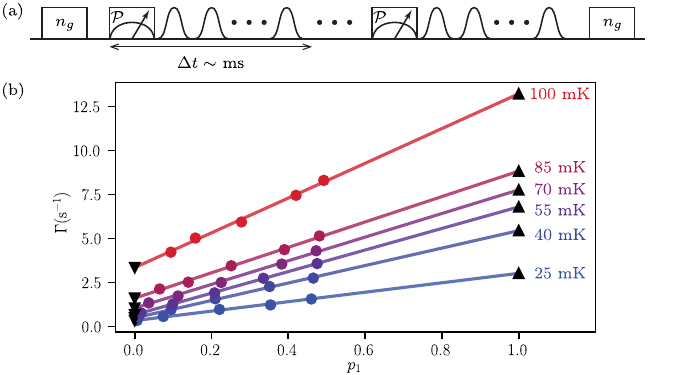}
\caption{Measurement of parity-switching rates at different qubit temperatures. This probes the effect of gap difference on tunneling of resident QPs occupying our films.
(a) Pulse sequence for the measurement of parity-switching rate conditioned on the qubit state, $\Gamma_0$ and $\Gamma_1$. Parity measurement is performed once every few milliseconds. Between the parity measurements, we perform a qubit scrambling pulse of a variable amplitude to manipulate the steady-state excited state population. To maintain a consistent offset charge $n_g$, we perform a Ramsey experiment to measure $n_g$, roughly once every 30 seconds (with some variation in this number across different measurements). Any observed drift is compensated by adjusting the gate voltage applied to the feedline, see Fig.~\ref{fig:wiring}.
(b) Parity-switching rate $\Gamma$ as a function of steady-state excited qubit state population $p_1$ at different temperatures. The solid line represents a linear fit of $\Gamma$ as a function of $p_1$, from which $\Gamma_0$ (down-pointing triangle) and $\Gamma_1$ (up-pointing triangle) can be extrapolated. The temperature corresponding to each measurement is indicated near the respective lines.
}\label{fig:sfig3}
  \end{center}
\end{figure*}
\subsection{Protocol of parity-switching measurement used to probe gap difference (Fig.~1 of the main text)}\label{sec:dD protocol}
In order to experimentally determine the gap difference at the Josephson junctions of our transmons, we measure their parity-switching rates at different fridge temperatures \cite{connolly_coexistence_2024}. These parity-switching rates have a temperature-dependent contribution which stems from the tunneling of resident QPs across the junction. The sensitivity of this contribution to the gap difference $\dD$ (see Section \ref{sec:dD theory}) allows us to measure $\dD$ with a high precision, as we describe below.

Fig.~\ref{fig:sfig3}(a) shows the pulse sequence used to measure the parity-switching rate. We monitor the charge parity of each device every few milliseconds for several minutes. The resulting parity jump trace is then fit to a hidden Markov model \cite{connolly_coexistence_2024} which can extract the parity-switching rate $\Gamma$ (even in the presence of measurement errors). Our ability to measure parity-switching events is highly sensitive to variations in the transmon offset charge $n_g$. This offset charge drifts randomly on the timescale of minutes. Therefore, we actively stabilize $n_g$ to perform parity-switching measurements. To this end, we measure $n_g$ every $\sim 10$ s \cite{serniak_direct_2019}, and based on this information we adjust $n_g$ by applying a DC voltage bias to the feedline \cite{diamond_distinguishing_2022}. If a significant drift occurs during a parity jump trace between two subsequent $n_g$ measurements, we discard this trace from our analysis.

As discussed in the main text, we measure the parity-switching rate separately for the qubit in the ground state $\ket{0}$ and the excited state $\ket{1}$. 
To this end, we measure the parity-switching rate while varying the steady-state population of the excited state, $p_1$. Specifically, we apply microwave pulses to scramble the qubit state between each parity measurement. An example of the measured $\Gamma$ as a function of $p_1$ is shown in Fig.~\ref{fig:sfig3}(b). The solid line represents a linear fit, which allows us to extrapolate $\Gamma_0=\Gamma\left(p_1=0\right)$ and $\Gamma_1=\Gamma\left(p_1=1\right)$.

Finally, we repeat the above measurements at different fridge temperatures, waiting at least 30 minutes after each change. This waiting period is needed for the temperature of the base plate of the dilution refrigerator to settle and also for the devices to thermalize with it. 

\begin{figure*}[h]
  \begin{center}
    \includegraphics[scale = 1]{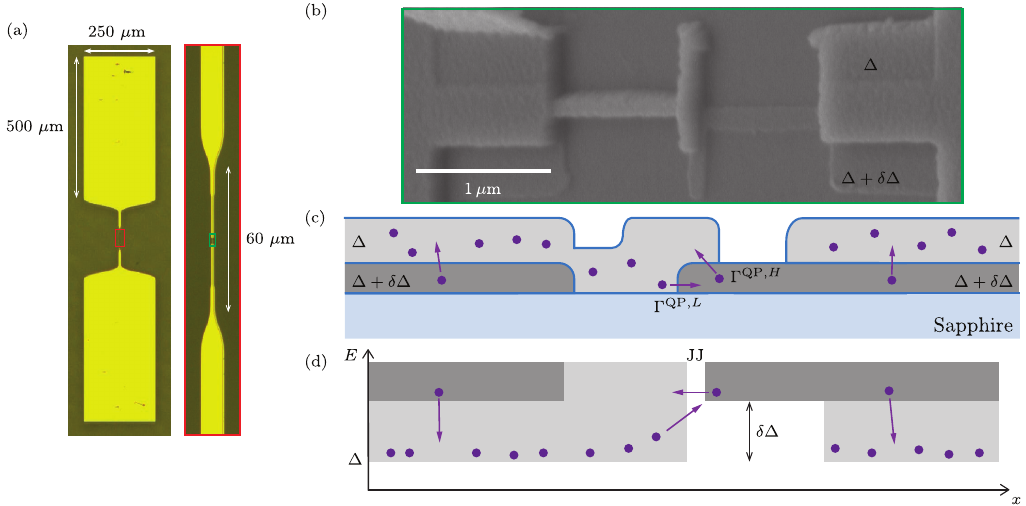}
\caption{ 
(a) Optical microscope image of a transmon qubit. The left panel shows the overall design and dimensions of the transmon in our chip. The right panel displays a magnified view of the red box on the left panel, focusing on the area near the junction. The green box indicates the location of the bridge-free Josephson junction.
(b) Scanning-electron-microscope (SEM) image of the bridge-free Josephson junction. The junction is formed at the overlap region of two distinct aluminum films. We intentionally reduce the thickness of the bottom film to create the difference in superconducting gaps at the junction.
(c) Schematic of the junction where the films are poisoned by QPs (purple circles). The superconducting gap in the thinner film is higher than that in the thicker film by an amount of $\dD$. As a result of the double-angle deposition process, two additional films, which do not contribute to the Josephson junction, are deposited in the pads on either side. Within each pad, most QPs reside in the low-gap film at small temperatures.
(d) Energy ``band'' diagram in the vicinity of the junction. The dark (light) gray regions correspond to the low-gap (high-gap) films on each side, while the white region is the gap. When the temperature of QPs is sufficiently low, the gap difference prevents the QPs from crossing the junction in both directions. First, the tunneling from left to right is suppressed by the zero density of states below $\Delta+\dD$ on the right of the junction. Second, most QPs on the right side are trapped in the low-gap film. Since this film is disconnected from the junction, this prevents QPs tunneling from right to left.
}\label{fig:sfig4}
  \end{center}
\end{figure*}

\subsection{Theoretical model for quasiparticle tunneling in the presence of gap difference}\label{sec:dD theory}
In this section, we develop the theory describing the tunneling of superconducting Bogoliubov quasiparticles across the Josephson junctions of our transmons.  Our transmons are gap-engineered, which means that there is a difference in superconducting gaps $\dD$ between the banks of the Josephson junction. In our theory, we account for this gap difference, following the approach of Refs.~\cite{diamond_distinguishing_2022, connolly_coexistence_2024,diamond_quasiparticles_2024}. The effects of the gap difference depend on whether the qubit frequency is larger or smaller than $\dD / h$. We consider both cases here. In our derivations, we assume that no phase bias is applied to the junctions. This is the situation relevant for our experiment.

We begin the exposition by describing the superconducting gap profile near the Josephson junction of our devices. The gap is controlled using the thickness of the films comprising the junction \cite{diamond_distinguishing_2022,marchegiani_quasiparticles_2022,connolly_coexistence_2024}. As shown in Fig.~\ref{fig:sfig4}, on the left of the junction, the film is thick, which leads to a lower value of the gap, $\Delta_L = \Delta$. On the right, the film is thin which leads to a higher value of the gap, $\Delta_H = \Delta+\dD$. Far away from the junction -- in the qubit pads -- both films are simultaneously present and the contact area between them is large. Thus, in the pads, QPs primarily reside in the low-gap film, as long as their energy is sufficiently small.

Throughout this section, we assume that the distribution function of the non-equilibrium QPs has Maxwell-Boltzmann form with a certain temperature $T$ and effective chemical potential $\mu$ measured from the Fermi level, $\mathcal{F}(\epsilon)\simeq\mathrm{exp}[-(\epsilon-\mu)/\kb T]$.
The temperature is assumed to be identical to that of the phonon bath in the substrate \cite{connolly_coexistence_2024}. This is justified if the QPs thermalize with the phonons sufficiently quickly (see Ref.~\cite{glazman_bogoliubov_2021} for timescales for such an equilibration). As described below, the chemical potential encodes the elevated density of the non-equilibrium QPs.
\subsubsection{Fermi's golden rule for the QP tunneling rates}

There are many different possible processes in which a single QP can tunnel across the Josephson junction of a transmon qubit. First of all, the QP can tunnel from the low-gap side of the junction to the high-gap side or vice versa. Second, the qubit state also might change during QP tunneling. If the qubit is initialized in state $\ket{i}$ (where $i=0,1,2...$ labels transmon states ordered according to their energy), after the QP tunneling it can transition to state $\ket{j}$. The energy difference between the qubit states is then absorbed by the tunneling QP. We thus characterize the QP tunneling by the partial rates $\Gamma_{ij}^{\mathrm{QP},W}$. Here $W=L,H$ describes whether the QP is initially on the low-gap side of the junction or on the high-gap side. Label $i$ indicates the initial qubit state and label $j$ indicates the final qubit state.

The QP tunneling rate for a qubit state transition ($\ket{i}\rightarrow \ket{j}$) from the low-gap side to high-gap side ($L\rightarrow H$) can be derived using Fermi’s golden rule:
\begin{equation}\label{eq:fgr}
\Gamma_{ij}^{\mathrm{QP},L}=\frac{16E_J}{h}\left(S_-^L[f_{ij}]\left|\bra{j}\cos\frac{\hat{\varphi}}{2}\ket{i}\right|^2 + S_+^L[f_{ij}]\left|\bra{j}\sin\frac{\hat{\varphi}}{2}\ket{i}\right|^2\right).
\end{equation}
The structure factors, $S_\pm^L$, account for constructive (+) or destructive (-) particle-hole interference during the QP tunneling. These factors can be calculated as:
\begin{equation}
S_\pm^L[f_{ij}] = \frac{1}{\bar{\Delta}}\int_{\epsilon_L>\Delta_L,\epsilon_H>\Delta_H} d\epsilon_L d\epsilon_H \frac{\epsilon_L\epsilon_H \pm\Delta_L\Delta_H}{\sqrt{\epsilon_L^2-\Delta_L^2}\sqrt{\epsilon_H^2-\Delta_H^2}}\,\mathcal{F}_L(\epsilon_L)\,\delta(\epsilon_H-\epsilon_L+hf_{ij}),
\end{equation}
where $f_{ij}\coloneqq f_j - f_i$ is the energy absorbed by the qubit from the tunneling QP ($hf_k$ is the energy of a transmon state $k$). Parameter $\bar{\Delta}=\Delta+\dD/2$ describes the superconducting gap averaged between the two sides of the junction.  $\mathcal{F}_{L,H}(\epsilon)$ describe the distribution functions for QPs in the low-gap and high-gap films, respectively. The non-equilibrium QP density -- reflecting either the QPs ``resident'' in the leads or QPs generated by high-energy impacts -- is reflected in the nonzero values of the chemical potentials. The relation between the chemical potentials $\mu_{L,H}$ and the QP density $\xqp^{L,H}$ (in units of the Cooper pair density) is \cite{glazman_bogoliubov_2021}:
\begin{equation}\label{eq:xqpLH}
    \xqp^{L,H}\coloneqq\frac{n_\mathrm{QP}^{L,H}}{2\nu_0\Delta_{L,H}}=\sqrt{\frac{2\pi\kb T}{\Delta_{L,H}}}\mathrm{e}^{-(\Delta_{L,H}-\mu_{L,H})/\kb T},
\end{equation}
where $\nu_0$ is the normal-state density of states at the Fermi level, and $n_\mathrm{QP}^{L,H}$ are the concentrations of the QPs in the low-gap and high-gap films, respectively.
This leads to
\begin{equation}
    S_\pm^L[f_{ij}] = \frac{1}{\bar{\Delta}}\int_{\epsilon_L>\max{(\Delta_L,\Delta_H+hf_{ij})}} d\epsilon_L \frac{\epsilon_L(\epsilon_L-hf_{ij}) \pm\Delta_L\Delta_H}{\sqrt{\epsilon_L^2-\Delta_L^2}\sqrt{(\epsilon_L-hf_{ij})^2-\Delta_H^2}}\,\mathrm{e}^{-(\epsilon_L-\mu_L)/\kb T}.
\end{equation}
By changing the variable to $z=\epsilon_L-\Delta-\alpha_{ij}$, where $\alpha_{ij}\coloneqq (\dD+hf_{ij})/2$, the structure factors become:
\begin{align}
\begin{split}
    S_\pm^L[f_{ij}] &= \frac{1}{\bar{\Delta}}\int_{\max{(-\alpha_{ij},\alpha_{ij})}}^\infty dz \frac{(z+\Delta+\alpha_{ij})(z+\Delta+\alpha_{ij}-hf_{ij})\pm\Delta(\Delta+\dD)}{\sqrt{(z+\Delta+\alpha_{ij})^2-\Delta^2}\sqrt{(z+\Delta+\alpha_{ij}-hf_{ij})^2-(\Delta+\dD)^2}}\,\mathrm{e}^{-z/\kb T}\mathrm{e}^{-(\Delta-\mu_L+\alpha_{ij})/\kb T}\\
    &= \xqp^L\sqrt{\frac{\Delta}{2\pi\kb T}}\frac{\mathrm{e}^{-\alpha_{ij}/\kb T}}{\bar{\Delta}}\int_{|\alpha_{ij}|}^\infty dz \frac{(z+\bar{\Delta}+hf_{ij}/2)(z+\bar{\Delta}-hf_{ij}/2)\pm(\bar{\Delta}-\dD/2)(\bar{\Delta}+\dD/2)}{\sqrt{z^2-\alpha_{ij}^2}\sqrt{(z+\alpha_{ij}+2\bar{\Delta}-\dD)(z-\alpha_{ij}+2\bar{\Delta}+\dD)}}\,\mathrm{e}^{-z/\kb T}.
\end{split}
\end{align}

We retain only the leading-order in $\dD/\bar{\Delta}$, $hf_{ij}/\bar{\Delta}$, and $\alpha_{ij}/\bar{\Delta}$ terms. This yields
\begin{align}\label{eq:sleq2}
\begin{split}
    S_+^L[f_{ij}] \simeq & \xqp^L\sqrt{\frac{\Delta}{2\pi\kb T}}\frac{\mathrm{e}^{-\alpha_{ij}/\kb T}}{\bar{\Delta}}\int_{|\alpha_{ij}|}^\infty dz \frac{2\bar{\Delta}^2}{\sqrt{z^2-\alpha_{ij}^2}\ \cdot2\bar{\Delta}}\,\mathrm{e}^{-z/\kb T}\\
    =&\xqp^L\sqrt{\frac{\Delta}{2\pi\kb T}}K_0\left(\frac{|\dD+hf_{ij}|}{2\kb T}\right)\:\mathrm{e}^{-(\dD+hf_{ij})/2\kb T},\\
    S_-^L[f_{ij}] \simeq & \xqp^L\sqrt{\frac{\Delta}{2\pi\kb T}}\frac{\mathrm{e}^{-\alpha_{ij}/\kb T}}{\bar{\Delta}}\int_{|\alpha_{ij}|}^\infty dz \frac{\bar{\Delta}z}{\sqrt{z^2-\alpha_{ij}^2}\ \cdot2\bar{\Delta}}\,\mathrm{e}^{-z/\kb T}\\
    =&\xqp^L\sqrt{\frac{\Delta}{2\pi\kb T}}\frac{|\dD+hf_{ij}|}{2\bar{\Delta}}K_1\left(\frac{|\dD+hf_{ij}|}{2\kb T}\right)\:\mathrm{e}^{-(\dD+hf_{ij})/2\kb T},
\end{split}
\end{align}
where $K_\nu$ are modified Bessel functions of the second kind.
\footnote{
Here, the following properties of Bessel functions are used:
\begin{equation*}
    K_0(x)=\int_x^\infty dt\frac{\mathrm{e}^{-t}}{\sqrt{t^2-x^2}}, \;
    K_1(x)=\frac{1}{x}\int_x^\infty dt\frac{t\mathrm{e}^{-t}}{\sqrt{t^2-x^2}}.
\end{equation*}
}

Similarly, for tunneling from the high-gap side to the low-gap side, the structure factors are given by:
\begin{align}\label{eq:sheq}
\begin{split}
    S_+^H[f_{ij}] = &\xqp^H\sqrt{\frac{\Delta+\dD}{2\pi\kb T}}K_0\left(\frac{|\dD-hf_{ij}|}{2\kb T}\right)\mathrm{e}^{(\dD-hf_{ij})/2\kb T}\\
    S_-^H[f_{ij}] = &\xqp^H\sqrt{\frac{\Delta+\dD}{2\pi\kb T}}\frac{|\dD-hf_{ij}|}{2\bar{\Delta}}K_1\left(\frac{|\dD-hf_{ij}|}{2\kb T}\right)\:\mathrm{e}^{(\dD-hf_{ij})/2\kb T}.
\end{split}
\end{align}
 
Finally, assuming QPs in the two electrodes are in equilibrium with the same phonon bath, we have $\mu_L=\mu_H$ and $T_L=T_H=T$. Together with Eq.~\eqref{eq:xqpHL}, this determines the ratio between the two QP densities as
\begin{equation}\label{eq:xqpHL}
    \frac{\xqp^H}{\xqp^L}=\sqrt{\frac{\Delta}{\Delta+\dD}}\mathrm{e}^{-\dD/\kb T}.
\end{equation}
Substitution of Eq.~\eqref{eq:xqpHL} into Eq.~\eqref{eq:sheq} yields
\begin{align}
\begin{split}
    S_+^H[f_{ij}] \simeq &\xqp^L\sqrt{\frac{\Delta}{2\pi\kb T}}K_0\left(\frac{|\dD-hf_{ij}|}{2\kb T}\right)\:\mathrm{e}^{-(\dD+hf_{ij})/2\kb T},\\
    S_-^H[f_{ij}] \simeq &\xqp^L\sqrt{\frac{\Delta}{2\pi\kb T}}\frac{|\dD-hf_{ij}|}{2\bar{\Delta}}K_1\left(\frac{|\dD-hf_{ij}|}{2\kb T}\right)\:\mathrm{e}^{-(\dD+hf_{ij})/2\kb T}.
\end{split}
\end{align}

The matrix elements in Eq.~\eqref{eq:fgr} can be approximated as follows in the transmon limit ($E_J\gg E_C$) \cite{catelani_parity_2014}:
\begin{align}
    \begin{split}
        \left|\bra{j}\cos\frac{\hat{\varphi}}{2}\ket{i}\right|^2 &\simeq \delta_{i,j}, \\ \left|\bra{j}\sin\frac{\hat{\varphi}}{2}\ket{i}\right|^2 &\simeq \delta_{i\pm1,j}\max(i,j)\sqrt{\frac{E_C}{8E_J}}.
    \end{split}
\end{align}
In summary, the total QP tunneling rate in both directions, $\Gqp_{ij} = \Gamma_{ij}^{\mathrm{QP},L} + \Gamma_{ij}^{\mathrm{QP},H}$, can be expressed as
\begin{align}\label{eq:Gqp}
    \begin{split}
        \Gqp_{ii} &= \xqp^L\frac{16E_J}{h}\sqrt{\frac{\Delta}{2\pi\kb T}}\frac{\dD}{\bar{\Delta}}K_1\left(\frac{\dD}{2\kb T}\right)\:\mathrm{e}^{-\dD/2\kb T} \\ 
        \Gqp_{i,i+1} &= \xqp^L\frac{16E_J}{h}(i+1)\sqrt{\frac{E_C}{8E_J}}\sqrt{\frac{\Delta}{2\pi\kb T}}\left[
        K_0\left(\frac{|\dD+hf_{i,i+1}|}{2\kb T}\right) + K_0\left(\frac{|\dD-hf_{i,i+1}|}{2\kb T}\right) 
        \right]\:\mathrm{e}^{-(\dD+hf_{i,i+1})/2\kb T}\\ 
        \Gqp_{i,i-1} &= \xqp^L\frac{16E_J}{h}i\sqrt{\frac{E_C}{8E_J}}\sqrt{\frac{\Delta}{2\pi\kb T}}\left[
        K_0\left(\frac{|\dD-hf_{i-1,i}|}{2\kb T}\right) + K_0\left(\frac{|\dD+hf_{i-1,i}|}{2\kb T}\right) 
        \right]\:\mathrm{e}^{-(\dD-hf_{i-1,i})/2\kb T}.
    \end{split}
\end{align}
The transition rates with $j\neq i-1,i,i+1$ are much smaller than those in Eq.~\eqref{eq:Gqp} in the limit $E_J\gg E_C$. In Fig.~1 of the main text, we measure quantities $\Gamma_{0}^{\rm QP}$ and $\Gamma_{1}^{\rm QP}$. In terms of rates given by Eq.~\eqref{eq:Gqp} they are defined as
\begin{equation}
    \label{eq:full_vs_partial}
    \Gqp_0\coloneqq\Gqp_{00}+\Gqp_{01}, \quad \Gqp_1\coloneqq\Gqp_{10}+\Gqp_{11}+\Gqp_{12}.
\end{equation}

\subsubsection{Quasiparticle density and its temperature dependence}
As shown in Eq.~\eqref{eq:Gqp}, the QP tunneling rates $\Gqp_{ij}(T)$ are determined by the density of the QPs in the films, $x_{\rm QP}^L$. This density has two contributions. First, there are resident non-equilibrium QPs populating the films. We assume that the total number of these resident QPs does not change with temperature. We encode this contribution in a parameter
\begin{equation}
x_{\rm QP}^{\rm ne} \coloneqq \frac{N_{\rm QP}}{2 \nu_0 \Delta V_L},
\end{equation}
where $N_{\rm QP}$ is the total number of resident QPs in the device and $V_L$ is \textit{the total volume of the low-gap film}. Second, there is a contribution to the QP density coming from the equilibrium QPs, $\xqp^{\rm th}$. This contribution is strongly temperature-dependent with $\xqp^{\rm th}\propto e^{-\Delta/\kb T}$. Combining the two contributions -- and assuming that they are independent from each other -- we express the total QP density in the low-gap film as \cite{connolly_coexistence_2024}
\begin{equation}\label{eq:xqpL}
    x_{\rm QP}^{L} = \zeta(T) x_{\rm QP}^{\rm ne} + \sqrt{\frac{2\pi \kb T}{\Delta}} e^{-\Delta/\kb T},\quad \zeta(T) = \frac{1}{1+\frac{V_H}{V_L}\sqrt{\frac{\Delta+\dD}{\Delta}}\mathrm{e}^{-\dD/\kb T}}.
\end{equation}
Here, $V_H$ is the total volume of the high-gap film, and parameter $\zeta(T)$ describes the redistribution of resident QPs ($N_{\rm QP}$) between the low-gap film and the high-gap film of the device. If the temperature is low, $\kb T\ll \dD$, this parameter is close to unity. This means that all QPs reside in the low-gap film. In the opposite limit, $\kb T\gg \dD$, we find $\zeta(T) = V_L / (V_H + V_L)$ (where we additionally assume $\delta\Delta \ll \Delta$). This means that the density of QPs in the low-gap film is suppressed due to their escape into the high-gap film.

Equations $\eqref{eq:Gqp}$, $\eqref{eq:full_vs_partial}$, and \eqref{eq:xqpL} fully define the calculation of the parity-switching rates $\Gamma_0^{\rm QP}$ and $\Gamma_1^{\rm QP}$.

\subsubsection{Approximate expressions for the QP tunneling rates}
In this section, we derive simplified expressions for the QP tunneling rates $\Gamma_{ij}^{\rm QP}$ under three assumptions. First, we assume $\Delta\gg\kb T$, such that the density of thermal QPs can be neglected compared to $\xqpne$. Second, we assume $h f_q \gg \kb T$. Finally, we assume that the circuit is in the transmon limit $E_J \gg E_C$. 

Before presenting the simplified expressions for the tunneling rates, we note that not all of the rates in Eq.~\eqref{eq:Gqp} are independent. First of all, in the limit of $E_J \gg E_C$, we have $\Gamma_{00}^{\rm QP}=\Gamma_{11}^{\rm QP}$. Second, as given by the detailed balance relation, the excitation rate is linked to the relaxation rate, $\Gamma_{01}^{\rm QP} = \exp\left(-h f_q/\kb T\right) \Gamma_{10}^{\rm QP}$. Finally, when $E_J \gg E_C$ it is appropriate to approximate\footnote{This approximate relationship only works sufficiently far from the resonances $\dD/h = f_{01}$ and $\dD/h = f_{12}$. In the vicinity of these resonances, its important to account for the difference $|f_{12}-f_{01}|\approx E_C / h$.} $\Gamma_{12} = 2 \Gamma_{01}$ since frequency $f_{12}$ is close to $f_{01}$ in this case. Therefore, in what follows, we only present the expressions for rates $\Gamma_{10}^{\rm QP}$ and $\Gamma_{11}^{\rm QP}$.

As described in the main text, the QP-induced relaxation rate, $\Gamma_{10}^{\rm QP}$, strongly depends on the relationship between the gap difference $\dD$ and qubit energy $hf_q$. This dependence arises because the qubit can donate its energy to help QPs overcome the gap difference and cross the junction. For this reason, we separately consider three regimes based on the relative magnitude of $\dD$, $hf_q$, and $\kb T$. We find
\footnote{Here, the following asymptotic forms of Bessel functions are used: $K_0(x\rightarrow\infty)\approx K_1(x\rightarrow\infty)\approx\sqrt{\pi/2x}\:\mathrm{e}^{-x}$, $K_0(x\rightarrow 0)\approx \ln(2/x)$, and $K_1(x\rightarrow 0)\approx 1/x$.}
\begin{enumerate}
    \item {
    zero-$\dD$ limit ($\dD\ll\kb T\ll hf_q$): 
    \begin{equation}
        \Gqp_{10}\approx\sqrt{\frac{8\Delta f_q}{h}}\cdot \frac{\xqpne}{1+V_H/V_L},\quad
        \Gqp_{11}\approx\frac{16E_J}{h}\sqrt{\frac{2\kb T}{\pi\Delta}}\cdot \frac{\xqpne}{1+V_H/V_L}
    \end{equation}
    }
    \item{
    resonant-$\dD$ limit ($|\dD-hf_q|\ll\kb T$): 
    \begin{equation}
        \Gqp_{10}\approx f_q \sqrt{\frac{2\Delta}{\pi\kb T}}\ln\left(\frac{4\kb T}{|\dD-hf_q|}\right)\cdot \xqpne,\quad
        \Gqp_{11}\approx \frac{16E_J}{h}\sqrt{\frac{\dD}{2\Delta}}\mathrm{e}^{-\dD/\kb T}\cdot \xqpne
    \end{equation}
    }
    
    \item{
    big-$\dD$ limit ($\dD-hf_q\gg\kb T$): 
    \begin{equation}\label{eq:big-dD}
        \Gqp_{10}\approx f_q \sqrt{\frac{2\Delta}{\dD-hf_q}}\:\mathrm{e}^{-(\dD-hf_q)/\kb T} \cdot \xqpne,\quad
        \Gqp_{11}\approx \frac{16E_J}{h}\sqrt{\frac{\dD}{2\Delta}}\mathrm{e}^{-\dD/\kb T}\cdot \xqpne
    \end{equation}
    }
\end{enumerate}

\subsection{Fitting parity-switching rate vs. temperature data}\label{sec:dD fit}
As explained in Section~\ref{sec:dD protocol}, our measurement yields the temperature dependence of the total parity-switching rates $\Gamma_0$ and $\Gamma_1$ in the steady state, i.e., away from the QP bursts. Here, we explain how we use theory of Section~\ref{sec:dD theory} to fit the measured temperature dependence. The resulting fits are shown in Fig~1 of the main text.

To fit the temperature dependence, we assume that the parity-switching comes from the two sources. The first source is the tunneling of QPs across the Josephson junction. This is the process described in Section~\ref{sec:dD theory}. The second source is the absorption of stray infrared photons at the junction \cite{houzet_photon-assisted_2019}. Accordingly, we can express the total parity-switching rates in the two qubit states as
\begin{equation}
\label{eq:two_contributions}
\Gamma_{i}(T)=\Gqp_i(T)+\Gamma_{i}^\mathrm{ph},
\end{equation}
where $\Gqp_i\coloneqq\Sigma_j\Gqp_{ij}$ is the total QP tunneling rate when the qubit is in state $\ket{i}$. Partial rates $\Gqp_{ij}$ are given by Eqs.~\eqref{eq:Gqp}. The contribution of the QPs determines the temperature dependence of Eq.~\eqref{eq:two_contributions}. 
In Eq.~\eqref{eq:two_contributions}, $\Gamma_{i}^\mathrm{ph}$ is the rate of parity-switching due to photon absorption, which we assume to be temperature-independent \cite{connolly_coexistence_2024}. In general, this rate can also depend on the qubit state, which we indicate with a subscript ``$i$''. As described in Section~\ref{sec:shielding}, in our experiment we attempted to suppress $\Gamma_{i}^\mathrm{ph}$ with filtering and shielding to enhance the sensitivity to $\Gqp_i(T)$.

\begin{figure*}[t]
  \begin{center}
    \includegraphics[scale = 1]{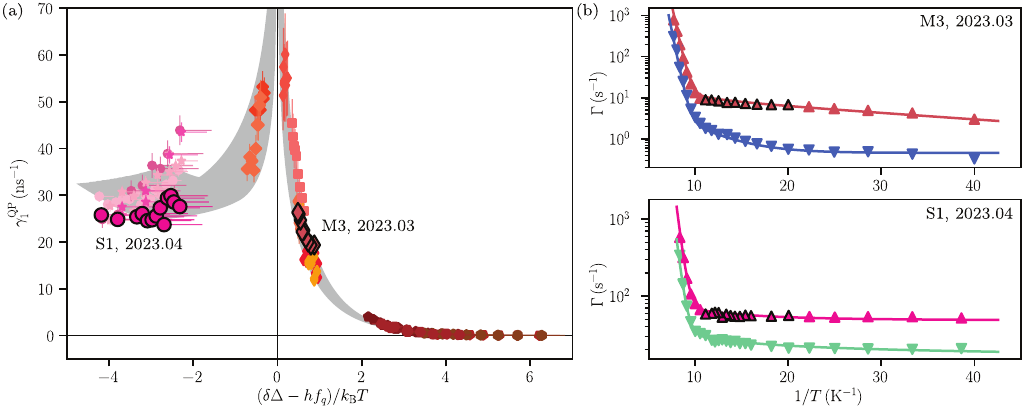}
\caption{ 
Panel (a) shows the dependence of QP tunneling rate on the gap difference, qubit frequency, and temperature. The vertical axis represents the QP tunneling rate when the qubit is in the excited state, ($\Gamma_1-\Gamma_1^\mathrm{ph}$), normalized by $\xqpne$: $\gqp_1=(\Gamma_1-\Gamma_1^\mathrm{ph})/\xqpne$. The horizontal axis shows a combination of parameters $\left[\dD-\hfq\right]/\kb T$ featured in Eq.~\eqref{eq:big-dD} for $\Gamma_{10}^{\rm QP}$. The parameters $\xqpne$ and $\Gamma_1^\mathrm{ph}$ entering the definition of $\gqp_1$ are obtained by fitting the temperature dependence of parity-switching rate; same is true for the parameter $\dD$ entering the definition of the x-axis.
The data in the plot are collected across various devices and cooldowns, indicated by the shape and color of markers, respectively. Makers sharing identical shape and color correspond to the data taken with the same device at different temperatures. To give an example, panel (b) shows full temperature sweeps for two devices, from which some of the data points in panel (a) are taken. Specifically, points with black outlines and negative values of the x-axis in panel (a) are taken from the bottom plot of panel (b). Points with black outlines and positive values of the x-axis are taken from the top plot of panel (b).
The error bars in panel (a) represent the fit uncertainties. The gray-shaded region illustrates the theoretical prediction.
}\label{fig:sfig5}
  \end{center}
\end{figure*}

Equations~\eqref{eq:Gqp}, \eqref{eq:xqpL}, and \eqref{eq:two_contributions} are used to fit the data for the temperature dependence of the parity-switching rates. The fits have five independent parameters: $x_\mathrm{QP}^{\rm ne}$, $\Delta$, $\dD$, $\Gamma^\mathrm{ph}_0$, $\Gamma^\mathrm{ph}_1$. Fig.~\ref{fig:sfig5}(b-c) show examples of the data and the corresponding fits for a medium-$\dD$ and small-$\dD$ devices, respectively.
The fit results for the all considered devices are summarized in Table~\ref{tab:stable}.

In Fig.~\ref{fig:sfig3}(a), we further illustrate the effect of $\dD$ on the parity-switching rate. To this end, we compare the measured parity-switching rates $\Gamma_1^{\rm QP}$ with theoretical predictions across all of the measured devices. Each device, as summarized in Table~\ref{tab:stable}, has distinct film thicknesses and thus a different $\dD$. The figure shows the parity-switching rates plotted against the ratio of the Arrhenius activation energy to the temperature, $\exponent$.  To account for variations in $\xqpne$ across devices and cooldowns, we normalize the parity-switching rates by $\xqpne$, $\gqp_1\coloneqq(\Gamma_1-\Gamma^\mathrm{ph}_1)/\xqpne$. Different marker shapes distinguish data from different devices, while color variations indicate different cooldowns for the same device. Each device is represented by multiple points in the plot corresponding to a temperature range $50\mathrm{mK} - 90\mathrm{mK}$. The gray shaded region represents the theoretical calculation for $\gqp_1$, as described by Eq.~\ref{eq:Gqp}. The theory prediction is a region -- as opposed to a single line -- because $\gqp_1$ is not solely determined by $\exponent$. Parameters such as $\delta\Delta$, $E_J$, $E_C$, $V_L$, and $V_H$ also explicitly enter the equations \eqref{eq:Gqp} for the rates. The shaded regions account for the variation of these parameters in our devices.

\begin{figure*}[h]
  \begin{center}
    \includegraphics[scale = 1]{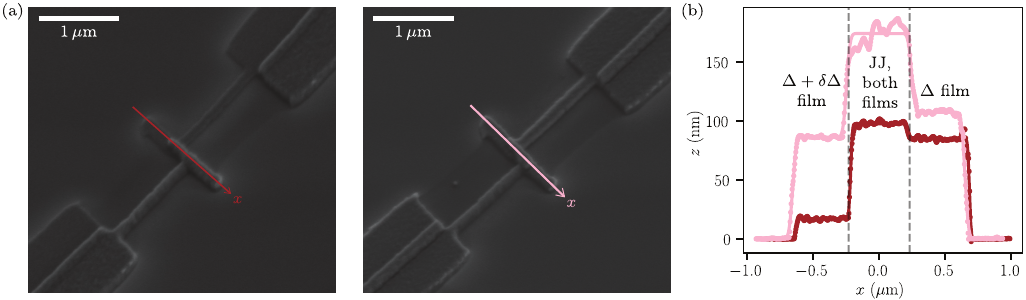}
\caption{
(a) SEM images of the Josephson junctions for the devices with ``big''-$\dD$ (left) and ``small''-$\dD$ (right) identical to the ones described in the main text. (b) Cross-section profiles from AFM measurements for the two devices featured in the main text. The thickness is measured along the line indicated in (a). Two distinct regions on either side correspond to the higher-gap film (left) and lower-gap film (right), while the central region represents the overlap of the two films. The thicknesses of the films are varied to control the superconducting gap difference at the junction.
}\label{fig:sfig6a}
  \end{center}
\end{figure*}

\begin{figure*}[h]
  \begin{center}
    \includegraphics[scale = 1]{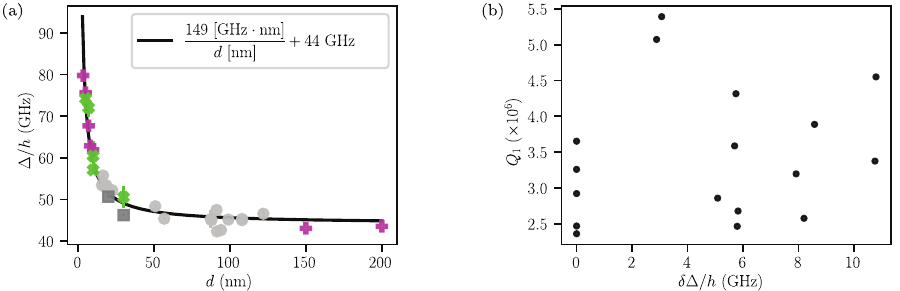}
\caption{
(a) Correlation between the superconducting gap of aluminum film, $\Delta$, and its thickness, $d$. The light gray dots represent the results of our measurements summarized in Table~\ref{tab:stable}. For each device, two data points, $\Delta$ and $\Delta+\dD$, are plotted. Data from previous measurements from the literature are shown by different marker shapes: purple + from Ref.~\cite{cherney_enhancement_1969}, green X from Ref.~\cite{court_energy_2007}, and gray squares from Ref.~\cite{connolly_coexistence_2024}.
(b) Transmon quality factor is roughly independent of the gap difference.
}\label{fig:sfig6b}
  \end{center}
\end{figure*}

\subsection{Correlation between the measured superconducting gaps and the film thicknesses}\label{sec:th_vs_D}
Precise control of the superconducting gap difference across the Josephson junction is the essence of the gap engineering. To achieve such control, we utilize a known trend that the superconducting gap of aluminum film increases with decreasing thickness. In this section, we demonstrate this trend in our devices using the values of $\Delta$ and $\delta\Delta$ extracted from parity-switching measurements. 

To investigate the relationship between the gap and the film thickness, we perform two steps.
First, we cool down our devices and measure the parity-switching rate as a function of temperature, as described in the previous sections. Fitting the data to the theory allows us to extract the superconducting gaps for the two films comprising the junction. Namely, the superconducting gap of the thicker film corresponds to parameter $\Delta$ of the fit, while the gap of the thinner film corresponds to the parameter $\Delta+\dD$. 
Second, we measure the thickness of the two films forming the junction using Atomic Force Microscopy (AFM). Fig.~\ref{fig:sfig6a}(b) shows cross-sectional AFM profiles of the big-$\dD$ and the small-$\dD$ devices discussed in the main text. Film thicknesses for other devices, similarly extracted from the AFM profile, are summarized in Table~\ref{tab:stable}.

The resulting superconducting gap versus film thickness, along with data from previous works \cite{chubov_dependence_1969,court_energy_2007,connolly_coexistence_2024}, is plotted in Fig.~\ref{fig:sfig6b}(a). The solid line represents the phenomenological fit \cite{marchegiani_quasiparticles_2022}, $\Delta/h=a/d+\Delta_0/h$, where $d$ is the film thickness, and $a$ and $\Delta_0$ are fit parameters.

Although the plot suggests that $\dD/h\gg 10$ GHz is achievable with sub-10 nm high-gap film, we do not deposit films thinner than 15 nm due to low yield in junction fabrication. We speculate that this low yield might be related to the surface roughness of our substrates. Nevertheless, for operational junctions, we do not observe a correlation between the film thickness and the qubit quality factor $Q_1\coloneqq 2\pi f_q T_1$ [see Fig.~\ref{fig:sfig6b}(b)]. We note that regardless of thickness, the contribution of resident QPs to the decoherence is negligible. These resident QPs can only be observed by measuring the parity-switching rates.

\clearpage
\DefTblrTemplate{firsthead, middlehead,lasthead}{default}{}
\begin{table}[hbt!]
\setlength{\belowcaptionskip}{0pt}
\caption{\label{tab:stable}
Experimental device specifications.
Film thicknesses for each device are measured using AFM (unless noted otherwise with superscripts). Parameter $E_J$, $E_C$, $f_q$, $T_1$ denote the Josephson energy, charging energy, qubit frequency, and qubit relaxation time, respectively.
The fit parameters -- $\dD$, $\Delta$, $\xqpne$, $\Gamma^\mathrm{ph}_0$, $\Gamma^{\rm ph}_1$ -- and their uncertainties are extracted from the temperature dependence of parity-switching rate, as outlined in Section~\ref{sec:dD fit}. 
The final column specifies the type of in-line IR filters used within the light-tight shield, see discussion in Section~\ref{sec:shielding}.
The devices measured together in a multiplexed package are indicated by the same row colors. The package material is either aluminum or copper. Copper package is used during cooldowns 2024.03, 2024.04.
The data presented in Fig.~1(a) of the main text was collected during cooldown 2024.04 (devices B4, M3, and S2). The burst measurement was performed with the same devices in a separate cooldown not listed in the table. The filtering configuration during this cooldown was identical to 2024.04.
}
\end{table}
\begin{longtblr}[
note{a} = {average value from measurements of other chips from the same wafer},
note{b} = {nominal value},
note{c} = {6cm CR-110 filter},
]
{
rowsep = 0.5pt,
colsep = 6pt,
vline{2-10} = {},
cells = {c},
cell{1}{1} = {r=2}{c},cell{1}{3} = {r=2}{c},
cell{1}{10} = {r=2}{c},
hline{6} = {-1pt},
hline{10} = {0pt},
cell{3}{1} = {r=2}{c},cell{3}{3-10} = {r=2}{c},
cell{5}{1} = {r=4}{c},cell{5}{3-10} = {r=2}{c,bg=gray9},
                      cell{7}{3-10} = {r=2}{c},
cell{9}{1} = {r=2}{c},cell{9}{3-10} = {r=2}{c,bg=gray8},
cell{11}{1} = {r=2}{c},cell{11}{3-10} = {r=2}{c,bg=brown8},
cell{13}{1} = {r=2}{c},cell{13}{3-9} = {r=2}{c},
cell{15}{1} = {r=4}{c},cell{15}{3-9} = {r=2}{c},
                       cell{17}{3-10} = {r=2}{c,bg=gray9},
cell{19}{1} = {r=10}{c},cell{19}{3-9} = {r=2}{c},
                       cell{21}{3-10} = {r=2}{c,bg=gray7},
                       cell{23}{3-9} = {r=2}{c},
                       cell{25}{3-10} = {r=2}{c,bg=brown9},
                       cell{27}{3-10} = {r=2}{c,bg=brown8},
cell{29}{1} = {r=6}{c},cell{29}{3-10} = {r=2}{c,bg=gray8},
                       cell{31}{3-10} = {r=2}{c,bg=gray7},
                       cell{33}{3-9} = {r=2}{c},
cell{35}{1} = {r=4}{c},cell{35}{3-10} = {r=2}{c,bg=brown9},
                       cell{37}{3-10} = {r=2}{c,bg=brown8},
}
\hline\hline
 device & film & cooldown & $E_J/h,\:E_C/h$ & $T_1$ & $\dD/h$ & $\Delta/h$ & $\xqpne $ & {$\Gamma_0^\mathrm{ph}$  ($\mathrm{s}^{-1}$)} & filter type \\* 
     & thickness    &  & $f_q$ (GHz)     & ($\mathrm{\mu s}$)   & (GHz)   &  (GHz)  & ($\times 10^{-10}$) &  {$\Gamma_1^\mathrm{ph}$  ($\mathrm{s}^{-1}$)}  &   \\*
  \hline
 B1 & 16 nm & 2022.10 & {5.89, 0.35 \\* 3.68} & 197 & 10.8$\pm$0.6 & 42.6$\pm$0.5 & 198$\pm$65 & {0.44$\pm$0.03 \\* 0.68$\pm$0.07
 } & {Eccosorb\\*CR-112}  \\* 
 & 94 nm & & & & & & &  & \\*  
 \hline 
 B2 & & 2022.12  & {7.26, 0.35 \\* 4.12} & 150 & 8.6$\pm$0.2 & 47.5$\pm$0.3 & 220$\pm$22 & {1.81$\pm$0.08\\*2.44$\pm$0.21
 }& {Eccosorb\\*CR-112}\\*  
  & 17 nm & & & & & & & & \\*  
 \cline{3-10}
  & 91 nm & 2023.01 & {7.22, 0.35 \\* 4.10} & 124 & 7.9$\pm$0.2 & 47.4$\pm$0.3 & 121$\pm$16 & {2.32$\pm$0.13 \\* 2.42$\pm$0.33}& {Eccosorb\\*CR-112}\\*  
  & & & & & & & & 
  & \\*  
 \hline
 B3 & 18 nm\TblrNote{a} & 2023.02 & {7.44, 0.36 \\* 4.23} & 127 & 10.8$\pm$0.6 & 42.3$\pm$0.6 & 137$\pm$49 & {0.50$\pm$0.04 \\* 0.72$\pm$0.09
 } &  {Eccosorb\\*CR-112} \\*  
  & 92 nm\TblrNote{a} & & & & & & & & \\*  
 \hline
 B4 & 19 nm & 2024.04 & {5.78, 0.34 \\* 3.57} & 115 & 8.2$\pm$0.4 & 45.2$\pm$0.4 & 6.3$\pm$1.8 & {0.22$\pm$0.01 \\* 0.40$\pm$0.02
 } & HERD1\\*  
  & 99 nm & & & & & & & & \\*  
 \hline
 M1 & 20 nm\TblrNote{b} & Ref~\cite{connolly_coexistence_2024} & {6.24, 0.36 \\* 3.83} & 193 & 4.5$\pm$0.1 & 46.2$\pm$0.3 & 5.3$\pm$0.6 & {0.13$\pm$0.02 \\* --}
 & Eccosorb\TblrNote{c}\\* 
  & 30 nm\TblrNote{b} & & & & & & & & CR-110\\*  
 \hline
 M2 & & 2022.11 & {5.70, 0.35 \\* 3.62} & 237 & 3.1$\pm$0.1 & 45.6$\pm$0.2 & 7.1$\pm$0.5 & {6.88$\pm$0.26 \\* --} &  Eccosorb\\*
  & 51 nm & & & & & & &&CR-112\\*  
 \cline{3-10}
  & 57 nm & 2022.12 & {5.75, 0.35 \\* 3.64} & 222 & 2.9$\pm$0.1 & 45.1$\pm$0.2 & 11.2$\pm$0.9 & {9.09$\pm$0.35 \\* --} & {Eccosorb\\*CR-112}\\*
  & & & & & & & & & \\*  
 \hline
 M3 & & 2023.03 & {9.77, 0.35 \\* 4.84} & 142 & 5.8$\pm$0.1 & 46.2$\pm$0.2 & 3.1$\pm$0.3 & {0.46$\pm$0.03 \\* --} & Eccosorb\\*
 & & & & & & & & & CR-112\\*  
 \cline{3-10}
  & & 2023.04 & {9.75, 0.35 \\* 4.84} & 94 & 5.1$\pm$0.1 & 47.9$\pm$0.2 & 12.8$\pm$1.9 & {7.72$\pm$0.44 \\* --} & {XMA\\*IR filter}\\*
  & & & & & & & & & \\*
  \cline{3-10}
  & 23 nm & 2023.06 & {9.77, 0.35 \\* 4.84} & 118 & 5.7$\pm$0.1 & 46.6$\pm$0.2 & 2.1$\pm$0.3 & {0.39$\pm$0.03 \\* --} & Eccosorb\\*
  & 122 nm & & & & & & & & CR-112\\*
  \cline{3-10}
  & & 2024.03 & {9.56, 0.35 \\* 4.79} & 89 & 5.8$\pm$0.1 & 46.1$\pm$0.1 & 1.5$\pm$0.1 & {0.28$\pm$0.01 \\* --} & HERD1\\*  
  & & & & & & & & & \\*
  \cline{3-10}
  & & 2024.04 & {9.55, 0.35 \\* 4.78} & 82 & 5.8$\pm$0.1 & 46.1$\pm$0.1 & 1.1$\pm$0.1 & {0.24$\pm$0.01 \\* --} & HERD1\\*  
  & & & & & & & & & \\*
 \hline
 S1 & & 2023.02 & {8.14, 0.35 \\* 4.40} & 118 & 0$\pm$0.4 & 45.3$\pm$0.4 & 16.9$\pm$0.7 & {4.23$\pm$0.46 \\* --} & {Eccosorb\\*CR-112}\\*
 & & & & & & & & & \\*
 \cline{3-10}
 & 87 nm\TblrNote{a} & 2023.04 & {7.98, 0.35 \\* 4.35} & 107 & 0$\pm$1.5 & 45.3$\pm$0.3 & 13.5$\pm$0.6 & {11.6$\pm$0.6 \\* --} & {XMA\\*IR filter} \\*
 & 108 nm\TblrNote{a} & & & & & & & & \\*
  \cline{3-10}
 & & 2023.05 & {7.90, 0.35 \\* 4.34} & 134 & 0$\pm$1.4 & 45.2$\pm$0.7 & 1.7$\pm$0.1 & {0.48$\pm$0.08 \\* --} & Eccosorb\\*
 & & & & & & & & & CR-112\\*
 \hline
 S2 & & 2024.03 & {7.69, 0.35 \\* 4.27} & 92 & 0$\pm$0.7 & 45.0$\pm$0.4 & 2.8$\pm$0.1 & {0.86$\pm$0.06 \\* --} & HERD1\\*  
 & 88 nm & & & & & & & & \\*
  \cline{3-10}
 & 108 nm & 2024.04 & {7.71, 0.35 \\* 4.27} & 88 & 0$\pm$1.1 & 45.1$\pm$0.4 & 1.9$\pm$0.1 & {0.52$\pm$0.05 \\* --} & HERD1\\*  
 & & & & & & & & & \\*
 \hline\hline
\end{longtblr}

\clearpage
\section{QP burst measurement and data analysis}
\label{sec:QP detect}
This section details the methodology used for characterizing QP burst events in our gap-engineered transmons. Section~\ref{sec:burst protocol} presents the protocol of the measurement that we use to investigate the bursts. Section~\ref{sec:burst analysis} outlines how we analyze the data collected with this protocol. This includes the detailed description of how we identify the burst events and how we filter out the false positives (potentially related to material defects in our transmons).
Section~\ref{sec:med-dD} includes the burst data similar to Fig.~2 of the main text but for the ``medium''-$\dD$ device [middle panel of Fig.~1(a), device M3 of Table~\ref{tab:stable}].
Section~\ref{sec:ng effect} demonstrates that most bursts in our device do not induce a significant shift in the offset charge $n_g$. Section~\ref{sec:parity burst} describes a way to detect weak bursts -- unresolvable by the measurement of Fig.~2 of the main text -- using the parity-switching measurements.
Finally, Section~\ref{sec:Tq msmt} elaborates on the measurement of the qubit temperature during burst events.

\subsection{Measurement protocol used for burst detection}\label{sec:burst protocol}
\begin{figure*}[h]
  \begin{center}
    \includegraphics[scale = 1]{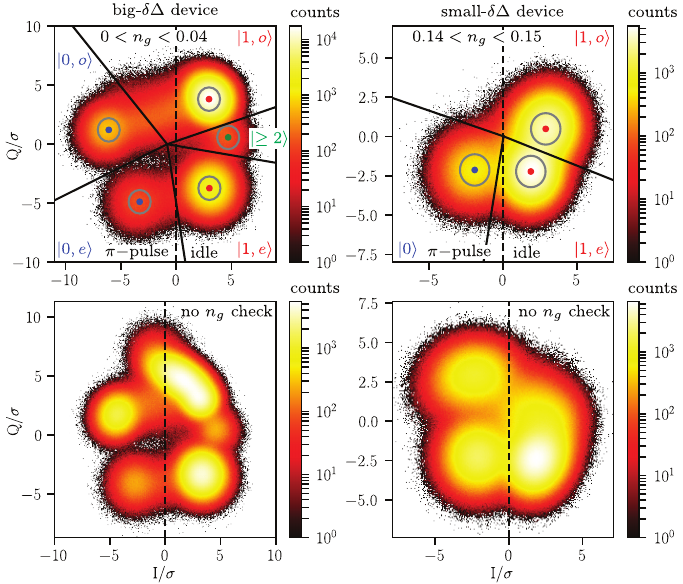}
\caption{ 
(Top) 2D histogram of readout outcomes for the devices in the main text (big- and small-$\dD$ devices). The qubits are measured every 5.7 $\mathrm{\mu s}$ for 90 seconds.
To continuously project the qubit into the excited state during the measurement, a $\pi$-pulse is applied immediately after each readout if the result is $I<0$. Solid lines show the thresholds used to differentiate the states $\ket{i,\mathcal{P}}$ from each other (here, $i\in\{0,1,2,\dots\}$ indicates the transmon level, and $\mathcal{P}\in\{o,e\}$ indicates the charge parity).
The offset charge $n_g$ is actively stabilized in a range indicated in the figure (in units of $2e$). To this end, before and after every 90 seconds of measurement, we probe $n_g$ with a Ramsey experiment \cite{diamond_distinguishing_2022}. We only retain the data if both measurements are in the required range. Otherwise the data is discarded and $n_g$ is corrected by adjusting the level of DC voltage applied to the feedline.
(Bottom) 2D histogram of discarded readout outcomes. Drifts in $n_g$ shift the position of the $\ket{1,o}$ state in the IQ plane, making it difficult to stabilize the qubit in its excited state using a fixed threshold ($I=0$).
}\label{fig:sfig7}
  \end{center}
\end{figure*}

We probe the QP bursts by repeatedly measuring the state of the qubit and applying feedback to keep it in the excited state $\ket{1}$. The idea is that if the qubit experiences an excess of the relaxation events, we can be reasonably confident that a burst is occurring. The data are collected in sequences of 90 s qubit readout traces (500 traces for big-$\dD$ device and 300 traces for small-$\dD$ device). As we now explain, this ``chopping'' of the data is needed to correct for the drift of the offset charge $n_g$ that happens on the minute timescale.

To explain why such a correction is needed, we note that our burst detection relies on the ability to accurately assign the readout outcomes to either state $\ket{0}$ or state $\ket{1}$. However, in our offset-charge-sensitive transmons this assignment is complicated by the $n_g$-sensitivity of the readout distributions (see next paragraph for explanation). Since $n_g$ drifts with time, this sensitivity makes it hard to consistently infer the transmon state for the single shot readout. To deal with this complication, we actively stabilize $n_g$ in a fixed narrow range. We achieve this by using Ramsey experiment to measure $n_g$ before and after each of the 90 s-long measurement sequences. If the offset charge drifts out of the desired range, we discard the measurement trace and apply the DC voltage to the feedline to correct it. 
Even with the fixed $n_g$, however, the readout distributions remain sensitive to parity, as shown in Fig.~\ref{fig:sfig7}. Nevertheless, we can still pick a threshold ($I=0$) which distinguishes between $\ket{0}$ and $\ket{1}$ states for both parities.

The $n_g$- and parity-sensitivity of the readout distributions warrants further explanation. This sensitivity results from that of the dispersive shifts of the readout resonator mode. Specifically, it stems from the strong hybridization of certain $n_g$-sensitive qubit state transitions [namely, $\ket{0}\rightarrow\ket{3}$ and/or $\ket{1}\rightarrow\ket{4}$] and the readout mode. Further details about this hybridization can be found in Refs.~\cite{serniak_direct_2019,connolly_coexistence_2024}.

Finally, a concern can be expressed regarding the effect of $n_g$ postselection on the burst detection rates that we report. We address this concern in Section~\ref{sec:ng effect}.

\subsection{Data analysis protocol and role of lossy two-level systems (TLSs)}\label{sec:burst analysis}
\begin{figure*}[h]
  \begin{center}
    \includegraphics[scale = 1]{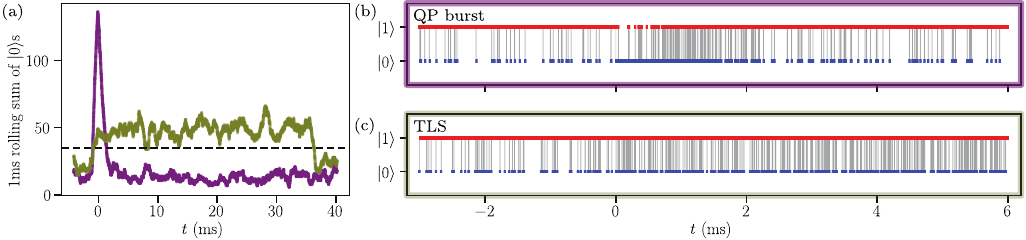}
\caption{
Results of the repeated qubit measurement, see Section~\ref{sec:burst protocol} for the protocol. The qubit is measured every $5.7\:\mu\mathrm{s}$ and is stabilized in the state $\ket{1}$. The stabilization is achieved by applying a $\pi$-pulse whenever the measurement outcome is $\ket{0}$. Counting the number of $\ket{0}$ outcomes in a 1 ms window, we identify two types of events: QP bursts and ``TLS'' events. (a) A 1 ms rolling sum of the number of ground state outcomes as a function of time during the QP burst (purple) and the ``TLS'' event (olive). The black dashed line indicates the burst threshold for the number of qubit relaxation events, defined in Fig.~2(c). Both of the shown events are above this threshold. However, their behavior is significantly different. When the qubit is coupled to a lossy TLS mode, the qubit relaxation rate is enhanced and sustained for tens of milliseconds. By contrast, a high-energy impact results in a sharp increase in the qubit relaxation rates, followed by fast decay.
(b-c) Readout trace zoomed-in around the time each event happened.
}\label{fig:sfig8}
  \end{center}
\end{figure*}

In this section, we describe the data analysis protocol used for burst identification. As we detail below, the identification is achieved by counting the number of measured qubit relaxation events in a certain time window (see Section~\ref{sec:burst protocol} for the measurement protocol). If this number is significantly elevated compared to its baseline value, we proclaim that a burst has occurred. There is, however, an important complication in this procedure. Namely, some of the detected events are not explained by either QP bursts or Poissonian statistical variation. Similarly to the bursts, in these events, the relaxation time of the qubit degrades. However, it does so much less severely than in the course of the burst events. Moreover, the duration of these events can exceed that of the bursts by more than an order of magnitude. In our interpretation, these events stem from the drift of qubit relaxation time which is related to changes in the dielectric environment. We explain below how we filter out such false positive events.

\subsubsection{Detecting the bursts}\label{sec:burst detection}
The search for burst begins by selecting ``burst candidates'’. This selection is performed using the procedure described in the main text. Namely, we break the readout traces into a sequence of 1 ms windows (containing 175 measurements each). 
We then count the number of relaxation events ($\ket{1}\rightarrow\ket{0}$) within each window. We only consider the process confined to the computational basis ($\ket{0}$, $\ket{1}$). This restriction can reduce false positive counts caused by the state leakage. 
The measurement accuracy does not allow to discriminate between the occupation of state $\ket{1}$ and higher excited 
states, meaning that even if the qubit is in a leakage state, the readout outcome would be a series of $\ket{1}$. This makes our detection protocol, which counts only number of $\ket{1}\rightarrow\ket{0}$ transitions, sensitive to the excess qubit relaxation rate but not to the leakage events. Additionally, we explicitly exclude transitions to or from the higher state that is distinguishable from the first excited state for the big-$\dD$ device (i.e., $\ket{\geq 2}$ in Fig.~\ref{fig:sfig7}). We note that population of this specific state remains below 1\% throughout our measurement. We discuss the effect of leakage on the measured relaxation rate in Section~\ref{sec:T1 analysis}.
If this number exceeds 35, the threshold shown in Fig.~2(c) of the main text, we mark the window as a ``burst candidate''. We note that the window duration is chosen to be close to the typical burst duration to maximize the sensitivity.

Subsequently, in the vicinity of each burst candidate, we locate the onset of the burst. To this end, we use the fact that right after the high-energy impact, the qubit relaxation rate surges. This means that the qubit stays in the ground state despite our attempts to bring it back to $\ket{1}$. For example, as shown in Fig.~2(d) and Fig.~\ref{fig:sfig8}(b), the relaxation rate can surpass the readout rate, leading to a sequence of consecutive $\ket{0}$ readouts immediately after the beginning of the burst. This allows us to define the onset of a burst event, $t=0$, as the point where the ground state population rapidly increases.

\subsubsection{False positives related to the TLSs}
However, not all of the ``burst candidate'' windows actually correspond to a QP burst generated by a high-energy impact. Namely, we find a fraction of candidates where the qubit relaxation rate is elevated consistently throughout the window. This behavior is different from a QP burst which leads to a transient spike in the relaxation rate. Moreover, the duration of the burst-unrelated events can vastly exceed 1 ms, i.e., the event can span multiple windows. The example of such an event is shown in Fig.~\ref{fig:sfig8}. As mentioned above, we interpret these events in terms of the drift of qubit relaxation time caused by changes in the dielectric environment. E.g., a lossy two-level system (TLS) can come in resonance with a qubit mode causing an elevated decay rate for a prolonged duration. This distinct temporal profile allows us to filter out such ``TLS'' events when looking for QP bursts.

\subsubsection{Time dependence of the relaxation rate during the burst}\label{sec:T1 analysis}

Now, having found the bursts and having identified $t=0$ for each of them, we average the signal over many bursts. In this way, we obtain the probability of a qubit relaxation event $p_{10}(t)$ as a function of time after the impact. 
This probability is evaluated from the data as $p_{10}(t)=(\textrm{num. of }\ket{1}\rightarrow\ket{0})/(\textrm{num. of }\ket{1})$; note that transitions to and from the detectable leakage states (such as $|\geq 2\rangle$ distribution in Fig.~\ref{fig:sfig7}) are by definition filtered out.
We then convert the probability into the qubit relaxation rate $\Gamma_{10}(t)$ as
\begin{equation}
	\Gamma_{10}(t)=\frac{-\ln(1-p_{10}(t))}{\Delta t}\ ,
\end{equation}
where $\Delta t=5.7\ \mathrm{\mu s}$ is the measurement cycle time. The resulting excess qubit relaxation rate, $\Delta\Gamma_{10}(t)\coloneq\Gamma_{10}(t)-1/T_1^{\rm steady}$, is shown in Fig.~2(e). Here, $T_1^{\rm steady}$ is the relaxation time of the qubit in the steady state, i.e., away from any QP bursts.

\subsubsection{Summary of the data analysis protocol}
To summarize, we analyze the measurement described in Section~\ref{sec:burst protocol} following the protocol below:
\begin{enumerate}
\item Burst candidate identification: Time intervals with more than 35 relaxation events within 1ms (= 175 readouts) are identified as burst candidates.
\item Determination of event onset: The starting point of each event is determined by locating the time at which the increase in the ground state population is the largest.
\item TLS event removal: We inspect burst traces individually and remove TLS-related events. 
\item Extraction of qubit relaxation rate during bursts.
\end{enumerate}

\subsection{QP bursts in the medium-$\dD$ device} \label{sec:med-dD}
\begin{figure*}[h]
  \begin{center}
    \includegraphics[scale = 1]{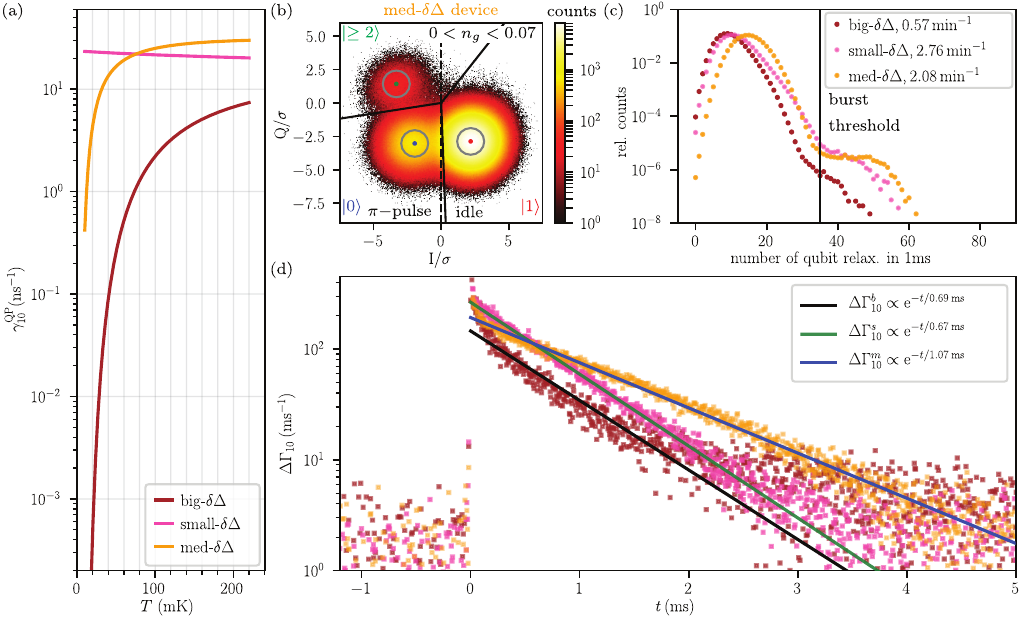}
\caption{QP bursts in the medium-$\dD$ device.
(a) The theoretically calculated dependence of the QP-induced relaxation rate $\gamma_{10}^{\rm QP}$ on temperature $T$ for three devices discussed in Fig.~1(a) of the main text. The rate is defined as $\gamma_{10}^{\rm QP} \coloneqq \Gamma_{10}^{\rm QP}/\xqp$. For the big-$\dD$ device with $\dD \gg hf_q$, the relaxation rate is exponentially suppressed at small temperatures. For the small-$\dD$ device, $\dD \ll hf_q$, the relaxation rate is roughly temperature-independent. For the medium-$\dD$ device, where $\dD$ mildly exceeds $hf_q$, the rate is also suppressed at small temperatures. However, at higher temperatures, $T\gtrsim 50\:\mathrm{mK}$, the relaxation rate in the medium-$\dD$ device becomes similar to that in the small-$\dD$ device. Since during the burst $T \sim 90\mK$, we expect the small- and medium-$\dD$ devices to have similar response to radiation impacts.
(b) 2D histogram of readout outcomes for the medium-$\dD$ device obtained from the burst measurement. The measurement protocol is similar to that described in Section~\ref{sec:burst protocol}. The dashed vertical line represents the threshold used to project the qubit state into the excited state after each readout. The offset charge, $n_g$, is confined to the narrow window.
(c) Histogram of the number of qubit relaxation events within a 1 ms window (analysis similar to that described in Section~\ref{sec:burst analysis}). In addition to data of Fig.~2(c) of the main text, we add the data from the medium-$\dD$ device (orange). For consistency, we use the same threshold to count QP burst events in all three devices. Notably, the number of detected bursts is \textit{similar} to that in the small-$\dD$ device and is significantly higher than that in the big-$\dD$ device.
(d) Excess qubit relaxation rate, $\Delta\Gamma_{10}=\Gamma_{10}-1/T_1^{\rm steady}$, as a function of time elapsed since high-energy impact. Here, $T_1^{\rm steady}$ is the steady-state relaxation time of the qubit away from the bursts. Unlike Fig.~2(e) in the main text, we present $\Delta\Gamma_{10}$ on a log scale to clearly illustrate the deviation from a single-exponential fit represented by solid lines.
}\label{fig:sfig9a}
  \end{center}
\end{figure*}

In the main text, we compare the QP bursts in two regimes distinguished by how $\dD$ compares to qubit energy. In the ``big'' gap difference regime, the gap difference substantially exceeds the qubit energy, $\dD\gg hf_q$. This suppresses the QP-induced decoherence during the bursts, as explained in Section~\ref{sec:dD theory}. On contrary, in the ``small'' gap difference regime, $\dD \ll hf_q$, the QP flow through the Josephson junction is not impeded. QPs are free to absorb the qubit energy and cause the decoherence of the qubit. In this section, we show that to suppress the effect of bursts, it is crucial to increase the gap difference well above the qubit energy. To this end, we investigate a device in an intermediate case where the gap difference is close to the qubit energy. We show that in this ``medium''-$\dD$ device the detection rate of the QP bursts is close to that in the small-$\dD$ device. This behavior agrees well with our theoretical expectation.

In our medium-$\dD$ device (device M3 in Table~\ref{tab:stable}), $\dD/h = 5.8$ GHz and $f_q = 4.8$ GHz, which means that $(\dD - hf_q)/\kb \approx 50\:{\rm mK}$. As is shown in Fig.~2(e), during the burst, the QP temperature $\approx 90\:{\rm mK}$ (see Section~\ref{sec:Tq msmt} for the description of the temperature measurement). Therefore, according to the theory presented in Section~\ref{sec:dD theory}, we do not expect the suppression of the QP tunneling by the gap difference. The behavior of this device during the QP bursts is summarized in Fig.~\ref{fig:sfig9a}(d). As is clear from the figure, the effect of the bursts in this device is comparable to that in the small-$\dD$ device. 
We note that the of the burst is the longest for the medium-$\dD$ device. We speculate that this distinction is unrelated to $\dD$. The decay of QP population can be explained by QP trapping. A particular mechanism how trapping can arise is the presence of vortices in the superconducting films \cite{nsanzineza_trapping_2014,vool_non-poissonian_2014,wang_measurement_2014}. The vortex configuration may be different for different samples. This would explain the difference in the burst decay time. Systematic investigation of such variation is beyond the scope of our work.

\subsection{Independence of the burst detection rate from $n_g$ post-selection} \label{sec:ng effect}
\begin{figure*}[h]
  \begin{center}
    \includegraphics[scale = 1]{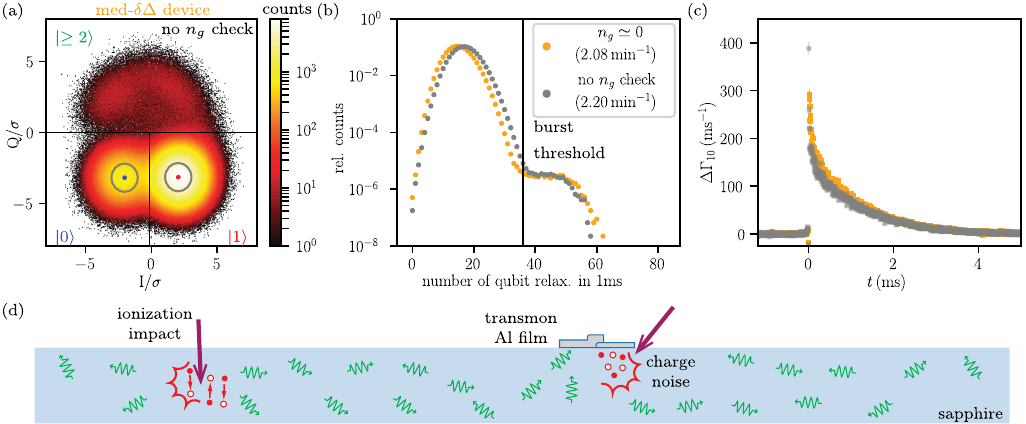}
\caption{
(a) 2D histogram of readout outcomes for the medium-$\dD$ device collected without $n_g$ postselection. Distributions corresponding to states $\ket{0}$ and $\ket{1}$ are offset-charge- and parity-independent. This allows us to perform the burst detection measurement outlined in Section~\ref{sec:burst protocol} without $n_g$ postselection.
(b-c) Comparison of results from two separate measurements, with and without $n_g$ postselection. Panel (b) presents the histogram of number of qubit relaxation events within a 1 ms window, and panel (c) shows the excess qubit relaxation rate after the high-energy impacts. Absence of post-selection increases the number of detected bursts by 6\%.
(d) When a high-energy particle collides with the substrate, it generates electron-hole pairs. These pairs recombine in a close vicinity of the impact site causing the local charge environment to change. The charge environment near the transmon is affected only when the impact occurs very close to the qubit island. In our design, the transmon pad area covers 0.2\% of the total chip area. Therefore, we expect only a small fraction of burst events to correlate with $n_g$ jumps, which indeed agrees with our observation.
}\label{fig:sfig9b}
  \end{center}
\end{figure*}
As described in Section~\ref{sec:burst protocol}, when measuring the bursts in the small- and big-$\dD$ devices, we have to rely on the offset charge $n_g$ postselection. Specifically, we only count the burst events where $n_g$ did not experience a significant change. However, it is known that the high-energy radiation impacts can significantly alter $n_g$ by redistributing the charges in the substrate \cite{wilen_correlated_2021}. 
This raises a concern about the reliability of our burst detection rates,
as our detection protocol may fail to detect bursts caused by ionizing impacts near the transmon island that induce significant charge noise, see Fig.~\ref{fig:sfig9b}(d).

In this section, we address this issue through a detailed investigation. We show that for our chips, $n_g$ postselection is justified and does not significantly skew the burst detection rates. To this end, we leverage measurements from the medium-$\dD$ device, where we can measure the burst detection rates both with and without $n_g$-postelection (which is allowed by the $n_g$-insensitivity of the readout distributions corresponding to states $\ket{0}$ and $\ket{1}$ in this device, see Fig.~\ref{fig:sfig9a}(b) and Fig.~\ref{fig:sfig9b}(a)).

\subsubsection{Burst detection rate with and without $n_g$ postselection}
To understand the effect of $n_g$ postselection, we repeat the measurement described in Section~\ref{sec:burst protocol} with and without $n_g$ check at the end of each trace. This modified measurement is allowed only for the medium-$\dD$ device, as its ``blob'' configuration for states $\ket{0}$ and $\ket{1}$ shows minimal drift with varying $n_g$, as shown in Fig.~\ref{fig:sfig9b}(a). This makes it possible to detect the bursts even if $n_g$ changed significantly within the 90 second measurement interval.
As is shown in Fig.~\ref{fig:sfig9b}(b), only $6\%$ more bursts are detected when we do not perform $n_g$ postselection. We therefore conclude that $n_g$ postselection is justified in our system (assuming that the big-$\dD$ and small-$\dD$ devices are similar in this regard).
Furthermore, we note that the time evolution of the excess qubit relaxation rate, $\Delta\Gamma_{10}(t)$, following the impact remains consistent regardless of whether the $n_g$ postselection is applied, as shown in Fig.~\ref{fig:sfig9b}(c).

\subsubsection{Reason why most bursts do not change $n_g$ in our devices}
Our results should be contrasted with a previous measurement of Ref.~\cite{diamond_distinguishing_2022}, where 
a different 3D aluminum transmon qubit on a sapphire substrate was measured. The authors found that $45\%$ of burst events were correlated with $n_g$ shifts. However, in the experiment of Ref.~\cite{diamond_distinguishing_2022} the chip was $\sim 3.6$ times smaller than in our work (while the transmon size was similar). Since the charge configuration only changes in a small range around the impact site of the high-energy particle, the impact was more likely to affect the offset charge of their transmon. In fact, if we account for the size difference, we expect $8\%$ of bursts to cause a significant $n_g$ shift based on measurements of Ref.~\cite{diamond_distinguishing_2022}. This is indeed close to the observed ratio of $6\%$.

\subsection{Detecting weak bursts by monitoring parity-switching} \label{sec:parity burst}
\begin{figure*}[h]
  \begin{center}
    \includegraphics[scale = 1]{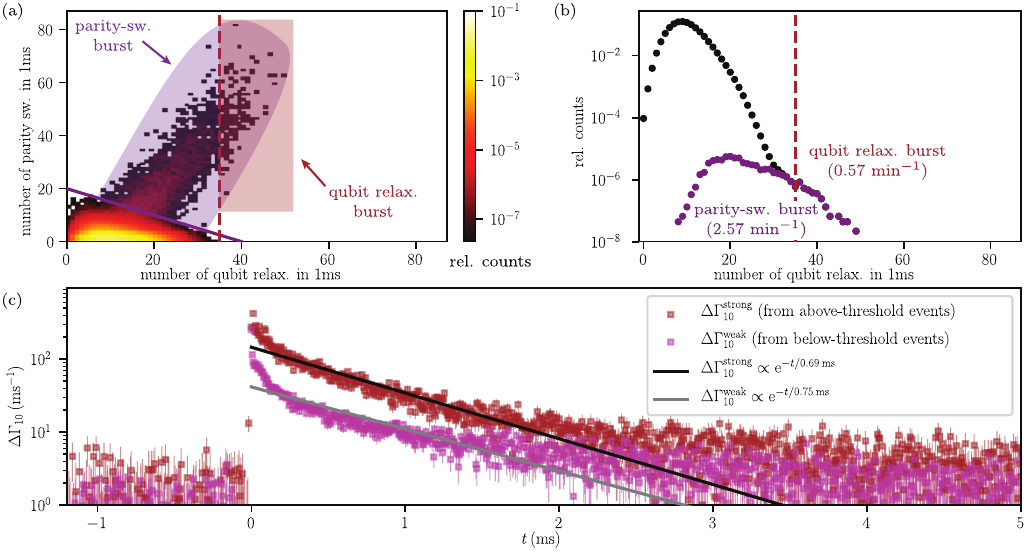}
\caption{
(a) Excess parity-switching events and excess qubit relaxation events in the big-$\dD$ device. 2D histogram visualizes the positive correlation between the two events. Similar to the 1D histogram in Fig.~2(c) of the main text, the histogram is described by a Poissonian process in the absence of QP bursts.
The effect of the high-energy impacts is represented by an anomalous diagonal tail (purple shaded region). We identify 1 ms windows where the two-dimensional threshold (purple solid line) is exceeded as ``parity-switching’’ burst events. The brown dashed line represents the threshold for the qubit relaxation burst events used in the main text. While this threshold is useful for comparing big-$\dD$ device to other devices (for which we do not have a parity measurement), it misses the majority of the high-energy impact events.
(b) 1D histogram projected onto the x-axis of panel (a). The black dots represent the total number of counts, which is identical to the brown data points in Fig.~2(c) of the main text. The purple dots represent the counts exceeding the 2D threshold in panel (a). The purple points to the left of the brown dashed line correspond to weaker bursts that are not caught by the burst detection protocol described in Section~\ref{sec:burst analysis}.
(c) Comparison of excess qubit relaxation rate, $\Delta\Gamma_{10}(t)$, between strong and weak bursts demonstrates similar time dynamics regardless of burst severity. We extract the relaxation rate for strong events, $\Delta\Gamma_{10}^\mathrm{strong}(t)$, by averaging the time profile of \textit{above}-threshold qubit relaxation bursts, see panel (b). We extract the relaxation rate for weak events, $\Delta\Gamma_{10}^\mathrm{weak}(t)$, by averaging the time profile of \textit{below}-threshold qubit relaxation bursts (detected via parity switching). 
The solid lines show single-exponential fits to the data.
}\label{fig:sfig10}
  \end{center}
\end{figure*}

In our big-$\dD$ device, the gap difference suppresses the influence of QPs on the qubit coherence. Therefore, by monitoring the qubit relaxation rate, we can only detect the strongest QP bursts. Weaker bursts are obscured by the steady-state qubit relaxation process, see histogram corresponding to the big-$\dD$ device in Fig.~2 of the main text. In this section, we explain how we can measure even these weaker bursts in our big-$\dD$ device. We achieve this by monitoring the parity-switching rates. Since the steady-state relaxation in our device is not caused by QP tunneling, it is not correlated with parity-switching. The QP bursts, on contrary, lead to a spike in the parity-switching rates. Therefore, QP bursts can be detected by monitoring parity of the transmon.

As we show in Fig.~\ref{fig:sfig7}, in the big-$\dD$ device, the single-shot readout provides additional parity information. This enables us to detect QP bursts by monitoring the parity-switching events.
Fig.~\ref{fig:sfig10}(a) shows a 2D histogram, where the x-axis represents the number of qubit relaxation events within a 1 ms window, and the y-axis represents the number of parity-switching events within a 1 ms window. 
A Poissonian process describes the primary distribution of qubit relaxation and parity-switching event counts. This process is characterized by a peak at $(1/T_1^{\rm steady},\Gamma)$, where $T_1^{\rm steady}$ is the qubit relaxation time and $\Gamma$ is the parity-switching rate in the steady state \footnote{In the absence of bursts, most of the parity switches detected in this measurement stem from readout assignment errors. The parity, in fact, mostly does not change in these events.}. In addition to the Poissonian peak, an anomalous diagonal tail is observed. This tail demonstrates a strong positive correlation between the excess qubit relaxation and the excess parity switches. We attribute such events to QP bursts given that QP tunneling simultaneously leads to parity-switching and qubit relaxation.

We define events exceeding the threshold marked by the purple angled line as ``parity-switching'' bursts, in contrast to the ``qubit relaxation'' bursts, which are identified only on excess qubit relaxation events (brown dashed line). This 2D thresholding allows us to capture more QP bursts, as shown in Fig.~\ref{fig:sfig10}(b). 
The panel (b) presents the histogram projected onto the qubit relaxation axis, showing that most parity-switching burst events occur below the $\ket{1}\rightarrow\ket{0}$ threshold (brown dashed line). Although these events are clearly caused by the high-energy impacts, their effect on the qubit state is weaker and indistinguishable from the steady-state qubit relaxation. Therefore, in the main text, only the top 20\% of the strongest bursts in the big-$\dD$ device are selected by thresholding based on excess qubit relaxation events. This criterion was necessary for consistent comparison across the three devices, as the joint parity-qubit statement measurement is not available in the other two devices.

Notably, the detected rate of parity-switching burst events (2.57 $\mathrm{min}^{-1}$) in the big-$\dD$ device is similar to that of the qubit relaxation burst events in the small-$\dD$ device (2.76 $\mathrm{min}^{-1}$). One possible interpretation is that the frequency of high-energy impacts is similar for the two devices, given that they share the same chip designs. However, the increase in QP density due to the impacts differently affects the qubit state, depending on the gap difference. Further experiments of parity-switching burst events are necessary to validate this hypothesis.

\subsection{Comparing time dynamics for strong and weak bursts}
The panel (c) compares the excess qubit relaxation rate, $\Delta\Gamma_{10}$, for two types of burst events. The brown points represent the strongest bursts above the threshold (i.e., identical to the data shown in Fig.2(e) and Fig.~\ref{fig:sfig9a}(d)). The purple dots show data for weaker parity-switching bursts, which fall on the left side of the brown dashed line in panel (b). 
We find that both events exhibit similar recovery dynamics after impacts, differing only in amplitude. Both show a non-exponential decay near the onset ($t\lesssim 100\mathrm{\mu s}$), but then return to the steady state roughly exponentially, with similar time constant of $\approx 0.7$ ms. 
This justifies our method for analyzing $\Delta\Gamma_{10}$, where we average the outcomes over all detected bursts regardless of their source. Although different types of ionizing radiation may cause distinct qubit responses, we at least confirm that the dynamics of $\Delta\Gamma_{10}$ are similar in shape for both more severe and weaker bursts.

\subsection{Elevated qubit temperature after high-energy impacts} \label{sec:Tq msmt}
In this section, we describe the measurement of qubit temperature during the QP bursts, see Fig.~2(e) of the main text. The measurement protocol is illustrated in Fig.~\ref{fig:sfig11}(a); it is conceptually similar to that of Section~\ref{sec:burst protocol}. Namely, the qubit is repeatedly measured while keeping $n_g$ within a narrow range. However, in contrast to the measurement of Section~\ref{sec:burst protocol}, we do not project the qubit into state $\ket{1}$. This allows us to detect the qubit excitation events. To increase the sensitivity of the measurement, we keep track of the parity information, as outlined in Section~\ref{sec:parity burst}. As such, the measurement of qubit temperature is only limited to the big-$\dD$ device, where the parity information can be obtained from the readout pulse. At the end of the section, we estimate the chip temperature using typical energy of high-energy particles causing the bursts and the known heat capacity of sapphire.

\subsubsection{Measurement of QP temperature after the bursts}
\begin{figure*}[h]
  \begin{center}
    \includegraphics[scale = 1]{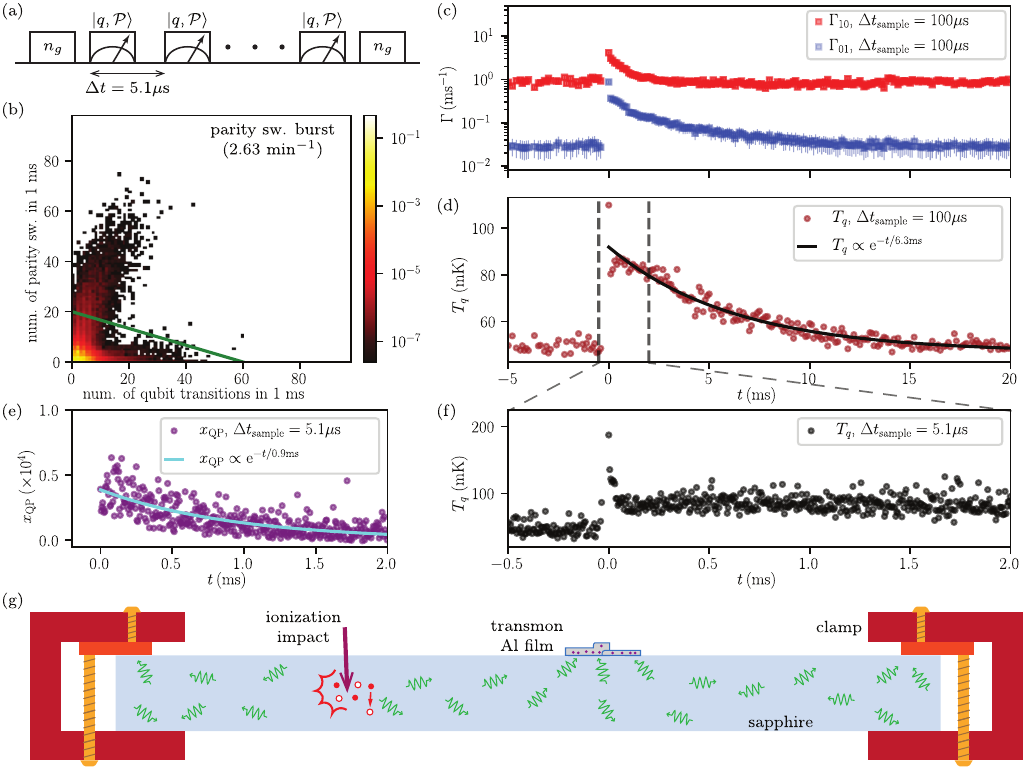}
\caption{ 
(a) Pulse sequence for the qubit temperature measurement in the big-$\dD$ device. We repeatedly readout the qubit every $5.1\mathrm{\mu s}$. Each readout pulse contains both the qubit state ($q \in \{0,1\})$ and the parity ($\mathcal{P} \in \{o,e\}$) information, see Fig.~\ref{fig:sfig7}. In contrast to measurement of Section~\ref{sec:burst protocol}, the qubit is not actively kept in the excited state. The offset charge $n_g$ is monitored every 90 s to ensure that no significant drift occurs during the measurement. 
(b) Search for parity-switching bursts. We identify the parity-switching bursts in our readout traces through the protocol outlined in the Section~\ref{sec:parity burst}. In contrast to Section~\ref{sec:parity burst}, now the x-axis of the histogram represents the number of qubit transitions in both directions ($\ket{1}\rightarrow\ket{0}$ and $\ket{0}\rightarrow\ket{1}$).
(c) Qubit relaxation rate, $\Gamma_{10}$, and excitation rate, $\Gamma_{01}$, as a function of time after the start of the burst event. All detected burst traces are aligned and averaged to extract $\Gamma_{10}(t)$ and $\Gamma_{01}(t)$. Each data point is averaged over a 100 $\mathrm{\mu s}$ interval.
(d) The qubit temperature, $T_q$, as a function of time after the onset of a burst event. The qubit temperature is calculated from $\Gamma_{10}$ and $\Gamma_{01}$ based on the detailed balance relation, Eq.~\ref{eq:detailed balance}. The solid line represents an exponential fit to the data, which indicates a recovery time of about 6.3 ms. 
(e) The time dynamics of QP density, $\xqp$, following a high-energy impact. $\xqp$ is calculated from $\Delta\Gamma_{10}$ and $\Delta\Gamma_{01}$, under the assumption that $\Tqp\approx T_q$ right after impacts, where QP tunneling events are dominant qubit decoherence channel. The time interval between data points ($\Delta t_\mathrm{sample}=5.1\mathrm{\mu s}$) is shorter here compared to panel (c-d). Notably, the $\xqp$ data is better described by the single exponential fit (solid line), unlike $T_q$ data in panel (d) and (f).
(f) The initial transient dynamics of the qubit temperature, $T_q$, captured by finer time resolution ($\Delta t_\mathrm{sample}=5.1\mathrm{\mu s}$). This data clearly reveals a rapid decay of $T_q$ from 200 mK within about 50 $\mathrm{\mu s}$, in addition to the long decay tail ($\approx 6$ ms) observed in panel (d). 
(g) Cross-sectional view of chip thermalization scheme. 
}\label{fig:sfig11}
  \end{center}
\end{figure*}

To identify QP bursts, we first plot the 2D histogram of measurement outcomes, see Fig.~\ref{fig:sfig11}(b). As the x-axis, we use the number of qubit transition ($\ket{1}\leftrightarrow\ket{0}$) events within a 1 ms window. As the y-axis, we use the number of parity-switching events, also within a 1 ms window. We then search burst events by applying the threshold marked by the green line. The identified rate of parity-switching bursts is 2.63 $\mathrm{min}^{-1}$, consistent with the measurement in Section \ref{sec:parity burst}.

Subsequently, we extract the qubit relaxation rate, $\Gamma_{10}$, and qubit excitation rate, $\Gamma_{01}$, as a function of time since the beginning of the burst. 
Similar to the procedure described in Section~\ref{sec:T1 analysis}, we first calculate the conditional probabilities, $p_{10}(t)$ and $p_{01}(t)$, and then convert them into the rates, $\Gamma_{10}(t)$ and $\Gamma_{01}(t)$, for each time interval. Fig.~\ref{fig:sfig11}(c) shows the results. Here, since we do not actively reset the qubit state, the measured $\ket{1}$ population is small under this protocol. This small population results in significant noise in the extracted $\Gamma_{10}(t)$, which we mitigate by averaging the data points over $\Delta t_\mathrm{sample}=100\:\mathrm{\mu s}$ window (in addition to averaging conditional probabilities over different bursts). 
We can then calculate the time dependence of the effective qubit temperature, $T_q$. This temperature is defined via the detailed balance relation,
\begin{equation}
\label{eq:detailed balance}
\Gamma_{10}/\Gamma_{01} = \exp{\left(\frac{\hfq}{\kb T_q}\right)}.
\end{equation}
Fig.~\ref{fig:sfig11}(d) shows the resulting $T_q(t)$. 

To observe the initial transient dynamics ($t \lesssim 100\:\mathrm{\mu s}$), we also plot the data for shorter time interval, $\Delta t_\mathrm{sample}=5.1\:\mathrm{\mu s}$, in panel (f).
We measure the effective qubit temperature as a proxy for the QP temperature, $\Tqp$, because the dominant decoherence mechanism -- for both $\Gamma_{10}$ and $\Gamma_{01}$ -- is QP tunneling during bursts. Assuming $\Tqp\approx T_q$, we can then extract the burst $\xqp$ using Eq.~\ref{eq:Gqp} as follows:
\begin{equation}
    \xqp=\frac{\Delta\Gamma_{10}+\Delta\Gamma_{01}}{\gqp_{10}(T\approx T_q)+\gqp_{01}(T\approx T_q)}
\end{equation}
The result is shown in Fig.~\ref{fig:sfig11}(e).

As shown in the earlier data in Fig.~\ref{fig:sfig11}(f), upon the onset of the QP burst, the qubit temperature abruptly jumps to $\approx 200$ mK but then quickly decays down to $\approx 90$ mK after about $50\:\mathrm{\mu s}$. We believe this spike is related to high-energy phonons that become down-converted rapidly. 
This is followed by more gradual return back to the steady state. 
The recovery timescale $\approx 6$ ms exceeds the duration of the burst in the qubit relaxation rate roughly by an order of magnitude. We attribute this prolonged recovery to the slow escape of hot phonons from our device, as described in the main text. Fig.~\ref{fig:sfig11}(g) illustrates the schematic of chip thermalization. The chip is clamped at both end with beryllium-copper springs, anchored by screws. The small point contacts at the clamps are the only places through which the phonons can escape the chip.

An important nuance in the measurement above is that even in the absence of bursts ($t<0$), the steady-state qubit temperature ($\approx 50$ mK) exceeds the base temperature of fridge ($\approx 20$ mK). This elevation is likely due to the readout-induced qubit transitions \cite{connolly_coexistence_2024}. These transitions stem from the strong hybridization between the readout tone and certain qubit state transitions, as described in Section~\ref{sec:burst protocol}.

\subsubsection{Estimate for chip heating due to impacts of high-energy particles}
When a high-energy particle -- such as a gamma-particle or a muon -- impacts our device, it deposits energy between 100 keV and 1 MeV into the chip \cite{fowler_spectroscopic_2024}. Some of that energy is absorbed into breaking Cooper pairs in the superconducting films forming our devices. However, a significant fraction of this energy ($\sim 50\%$) goes to phonons whose energy is insufficient to break the Cooper pairs. Thus, we expect the effective temperature of the chip to surge during the burst. In this section, we use Debye model to estimate the temperature of our chip after the impact of a high-energy particle.

As given by Debye model, the internal energy $U$ of our sapphire chip as a function of its temperature $T$ can be calculated as
\begin{equation}
    \label{eq:debye}
    U(T)=9N\kb\,\frac{T^4}
    {T_\mathrm{D}^3}\int_0^{T_D/T} dx \frac{x^3}{\mathrm{e}^x-1},
\end{equation}
where $N$ is the number of molecules, and $T_D\approx1000$ K is the Debye temperature of sapphire \cite{fugate_specific_1969,viswanathan_heat_1975}. For our 4 cm $\times$ 4 mm $\times$ 0.5 mm sapphire chips, we estimate the prefactor $9N\kb\approx 0.23$ J/K (accounting for the density of sapphire, its atomic mass, and the chip size).

As mentioned above, the ionizing radiations typically deposit energy on the scale of 100 keV to 1 MeV. For simplicity, let us assume that this energy is completely converted into the heating of the chip. Then, using Eq.~\eqref{eq:debye}, we find that the chip temperature after the burst should be in the range between 60 mK and 100 mK. This range aligns closely with our experimental observation.
This supports our hypothesis that the elevated qubit temperature after the high-energy impact can be attributed to the hot phonons stuck in the substrate.

\newpage
\bibliography{references.bib}